\documentclass[fleqn,usenatbib]{mnras}

\usepackage{newtxtext,newtxmath}
\usepackage[T1]{fontenc}

\DeclareRobustCommand{\VAN}[3]{#2}
\let\VANthebibliography\thebibliography
\def\thebibliography{\DeclareRobustCommand{\VAN}[3]{##3}\VANthebibliography}

\usepackage{graphicx}	% Including figure files
\usepackage{amsmath}	% Advanced maths commands
\usepackage[percent]{overpic}

\title[Fe, temperature and winds on WASP-76~b]{Spatially-resolving the terminator: Variation of Fe, temperature and winds in WASP-76~b across planetary limbs and orbital phase}
\author[S. Gandhi et al.]{
Siddharth Gandhi$^{1,2,3}$\thanks{E-mail: gandhi@strw.leidenuniv.nl},
Aurora Kesseli$^{1,4}$,
Ignas Snellen$^{1}$,
Matteo Brogi$^{2,5,3}$,
Joost P. Wardenier$^{6}$,
\newauthor
Vivien Parmentier$^{6}$,
Luis Welbanks$^{7}$\thanks{NHFP Sagan Fellow}
and Arjun B. Savel$^{8}$
\\
$^{1}$Leiden Observatory, Leiden University, Postbus 9513, 2300 RA Leiden, The Netherlands\\
$^{2}$Department of Physics, University of Warwick, Coventry CV4 7AL, UK\\
$^{3}$Centre for Exoplanets and Habitability, University of Warwick, Gibbet Hill Road, Coventry CV4 7AL, UK\\
$^{4}$IPAC, Mail Code 100-22, Caltech, 1200 E. California Blvd., Pasadena, CA 91125, USA\\
$^{5}$INAF-Osservatorio Astrofisico di Torino, Via Osservatorio 20, I-10025, Pino Torinese, Italy\\
$^{6}$Department of Physics (Atmospheric, Oceanic and Planetary Physics), University of Oxford, Oxford, OX1 3PU, UK\\
$^{7}$School of Earth \& Space Exploration, Arizona State University, Tempe, AZ 85257, USA \\
$^{8}$Department of Astronomy, University of Maryland, College Park, MD 20782, USA
}

\date{Accepted XXX. Received YYY; in original form ZZZ}

\pubyear{2022}

\begin{document}
\label{firstpage}
\pagerange{\pageref{firstpage}--\pageref{lastpage}}
\maketitle

\begin{abstract}
Exoplanet atmospheres are inherently three-dimensional systems in which thermal/chemical variation and winds can strongly influence spectra. Recently, the ultra-hot Jupiter WASP-76~b has shown evidence for condensation and asymmetric Fe absorption with time. However, it is currently unclear whether these asymmetries are driven by chemical or thermal differences between the two limbs, as precise constraints on variation in these have remained elusive due to the challenges of modelling these dynamics in a Bayesian framework. To address this we develop a new model, HyDRA-2D, capable of simultaneously retrieving morning and evening terminators with day-night winds. We explore variations in Fe, temperature profile, winds and opacity deck with limb and orbital phase using VLT/ESPRESSO observations of WASP-76~b. We find Fe is more prominent on the evening for the last quarter of the transit, with $\log(X_\mathrm{Fe}) = {-4.03}^{+0.28}_{-0.31}$, but the morning shows a lower abundance with a wider uncertainty, $\log(X_\mathrm{Fe}) = {-4.59}^{+0.85}_{-1.0}$, driven by degeneracy with the opacity deck and the stronger evening signal. We constrain 0.1~mbar temperatures ranging from $2950^{+111}_{-156}$~K to $2615^{+266}_{-275}$~K, with a trend of higher temperatures for the more irradiated atmospheric regions. We also constrain a day-night wind speed of $9.8^{+1.2}_{-1.1}$~km/s for the last quarter, higher than $5.9^{+1.5}_{-1.1}$~km/s for the first, in line with general circulation models. We find our new spatially- and phase-resolved treatment is statistically favoured by 4.9$\sigma$ over traditional 1D-retrievals, and thus demonstrate the power of such modelling for robust constraints with current and future facilities.
\end{abstract}


%%%%%%%%%%%%%%%%%%%%%%%%%%%%%%%%%%%%%%%%%%%%%%%%%%

%%%%%%%%%%%%%%%%% BODY OF PAPER %%%%%%%%%%%%%%%%%%

\section{Introduction}

Ultra-hot Jupiters (UHJs), with temperatures in excess of 2000~K, are ideal laboratories for exoplanet study given their hot extended atmospheres and large signal-to-noise. %They are also of particular relevance given that these strongly irradiated exoplanets have no such Solar System analogue, and the processes that formed them and moved them to their present location are not fully understood. 
The temperature for these exoplanets is high enough that refractory species remain gaseous and therefore detectable in the photosphere. These conditions are therefore ideal for studying study a range of physical phenomena at such extreme temperatures, such as the thermal dissociation, rain-out and ionisation of atomic species. In recent years, a number of UHJs have been characterised, with numerous species detected in their atmosphere in both primary and secondary eclipse using ground based high-resolution spectroscopy (HRS) at optical wavelengths \citep[e.g.,][]{hoeijmakers2019, seidel2019, bourrier2020, pino2020, cabot2020, merritt2021}. These observations are ideal to detect trace species in the atmosphere due to their high sensitivity and resolution of R$\gtrsim$25,000, allowing for clear absorption and emission lines to be extracted from the spectra \citep[see e.g., review by][]{birkby2018}.

WASP-76~b has had one of the most well-characterised atmospheres ultra-hot Jupiters to date with a mass of 0.894~$M_\mathrm{J}$, radius of 1.854~$R_\mathrm{J}$ and an equilibrium temperature of $\sim$2160~K with full redistribution \citep{west2016}. Recently, \citet{ehrenreich2020} showed variation in the strength and position of the Fe signal in WASP-76~b with primary eclipse observations using the ESPRESSO spectrograph on the VLT, and proposed that this variation could be a result of condensation of Fe on the night side of the atmosphere. These observations have also been used to detect numerous atomic and ionic species in the atmosphere \citep{tabernero2021, kesseli2022} and to resolve more of the asymmetric condensation features for some species \citep{kesseli2022}. The asymmetry and relative shift of $\sim$10~km/s in the position of the Fe signal during the transit was additionally confirmed by observations with HARPS \citep{kesseli2021}, which were also previously used to detect and study Na in the atmosphere \citep{seidel2019, zak2019}. Both of these observations have also been combined to constrain day-night and vertical winds in the atmosphere by studying the Na doublet \citep{seidel2021}.

A range of 3-dimensional (3D) general circulation models (GCMs) have been used to interpret the asymmetric Fe signal in the atmosphere of WASP-76~b. Such complex models simulate a wide range of physical and chemical processes and provide an unparalleled levels of spatial and dynamical detail for hot Jupiters \citep[e.g.,][]{showman2009, rauscher2013, dobbs-dixon2013, mayne2014}, as well as ultra-hot Jupiters \citep[e.g.,][]{tan2019}. Importantly, such 3D models can also be used to predict spatial variability in the abundance of the chemical species and temperature structure of the atmosphere and hence in each limb of the terminator, as well as velocity shifts of the spectra due to winds. These models have shown that the Fe asymmetry in WASP-76~b could be a result of cloud formation on the cooler morning side \citep{savel2022}, or the thermal asymmetry and wind profiles within the atmosphere \citep{wardenier2021}. Hence, it is currently unclear whether Fe is inherently weaker on the leading limb due to its lower abundance or whether clouds/temperature can affect the signal significantly to alter its strength.

In contrast to GCMs, retrieval frameworks of exoplanetary atmospheres determine the best fitting parameters for exoplanet atmospheres from observations, often exploring many millions of models and encompassing a wide range in temperature, clouds and abundances of relevant species. Such models generally assume spatially-homogeneous atmospheric conditions throughout the transit. Retrievals have been ubiquitous for interpretation of low-resolution observations in the last decade, most notably for HST spectra in both emission and transmission geometries \citep[e.g.,][]{madhu2009, madhu2014, line2016, barstow2017, mikal-evans2019, benneke2019, zhang2020}. However, their use in analysing high-resolution spectra has been relatively recent \citep[e.g.,][]{brogi2017, gandhi2019_hydrah, pelletier2021, line2021, gibson2022}, and has been made possible thanks to the development of cross-correlation to likelihood mappings \citep{brogi2019, gibson2020}, which allow for Bayesian analyses to be performed on the data. However, high-resolution retrievals remain a computational challenge due to the many millions of models required at high resolutions of R$\gtrsim$25,000 and the various analysis techniques used to extract the planetary signal from the noise.

Winds in exoplanetary atmospheres have also recently been detected by a range of ground based high-resolution observations, and result in an observable Doppler shift and/or broadening of the overall spectrum arising from the range in velocities along the line of sight. \citet{snellen2010} detected the winds and orbital motion of an exoplanet for the first time through high-resolution ground based observations in the infrared. \citet{louden2015} have constrained the eastward winds in the hot Jupiter HD189733~b through forward modelling of the Na absorption lines, showing that including such dynamics is key for accurate characterisation. Furthermore, GCMs have also been shown to provide better fits to high-resolution observations than 1D models which do not incorporate thermal variation and dynamical processes \citep{beltz2021}. Including winds into atmospheric models is also of particular importance given that there are also a growing number of high-resolution studies which have constrained these on the terminator \citep{brogi2016, flowers2019, seidel2021}. In addition, recent work has also shown that cloud formation on the night side of UHJs is strongly dependent on the atmospheric dynamics \citep{komacek2022} and these clouds can affect phase-dependent emission spectra.

Ultimately, what is needed is a model that is capable of interpreting the variability in the observations of a UHJ, such as WASP-76~b, whilst simultaneously incorporating dynamical processes in a statistical framework. We must include as much of the information content of a GCM whilst keeping the model computationally tractable with the minimum number of additional parameters to encompass the relevant physics. Such a retrieval model must therefore be capable of constraining separate Fe abundances, thermal profiles and opacity decks for the morning (leading) and evening (trailing) limbs, and include winds. The winds will act to not only transfer thermal energy and material between the day and night sides of the atmosphere, but also shift the spectrum for each limb according to the velocity profile.

Here we develop HyDRA-2D, a spatially-resolved retrieval model which incorporates separate constraints on morning and evening limbs and a wind travelling between the day and night sides of the atmosphere. Such a model represents the first step into multi-dimensional retrievals on high-resolution data. HyDRA-2D allows for a full exploration of the parameter space using Bayesian algorithms whilst capturing the dynamical effects which can strongly influence the observed spectra of exoplanets. We separate the atmosphere into a morning and evening limb, as these regions of the terminator exhibit different velocities due to the planet's rotation, as well as different thermochemical states. This rotation therefore results in separate spectral contributions for each side. 

We use HyDRA-2D to retrieve the two nights of ESPRESSO observations of WASP-76~b \citep{ehrenreich2020} and determine the variability in Fe, temperature, opacity and wind speed across each limb of the planet. Whilst there are additional transit observations of WASP-76~b observed with HARPS \citep{seidel2019, kesseli2021}, we restrict ourselves to the VLT observations due to the higher signal-to-noise ratio and spectral resolution available with ESPRESSO, which is vital for discerning the different regions of the planet that are separated by small velocity shifts. We perform two separate retrievals for the $\phi = -0.04 \,\text{-}\, -0.02$ and $\phi = 0.02 \,\text{-}\, 0.04$ phase ranges of the observations, with $\phi=0$ corresponding to the mid-transit time. The chosen phase ranges allow us to also explore the variation of atmospheric properties with phase, given that the region of the planetary atmosphere probed varies by $\sim30^\circ$ between ingress and egress \citep{wardenier2022}. We exclude the phases between $\phi = -0.02$ and $\phi = +0.02$ from our analysis owing to the Doppler shadow of the star crossing the planetary signal during mid-transit. In addition to retrievals with HyDRA-2D, we also perform a set of spatially-homogeneous 1D retrievals across the combined phase range as well as for each of the ranges separately to compare the differences in the constrained parameters with model sophistication.

In the next section we discuss our HyDRA-2D retrieval setup, followed by the results and discussion section where we discuss our constraints for each side of the atmosphere and comparisons with GCMs, and finally end with the concluding remarks.

\begin{figure*}
\centering
	\includegraphics[width=\textwidth,trim={0cm 0cm 0cm 0},clip]{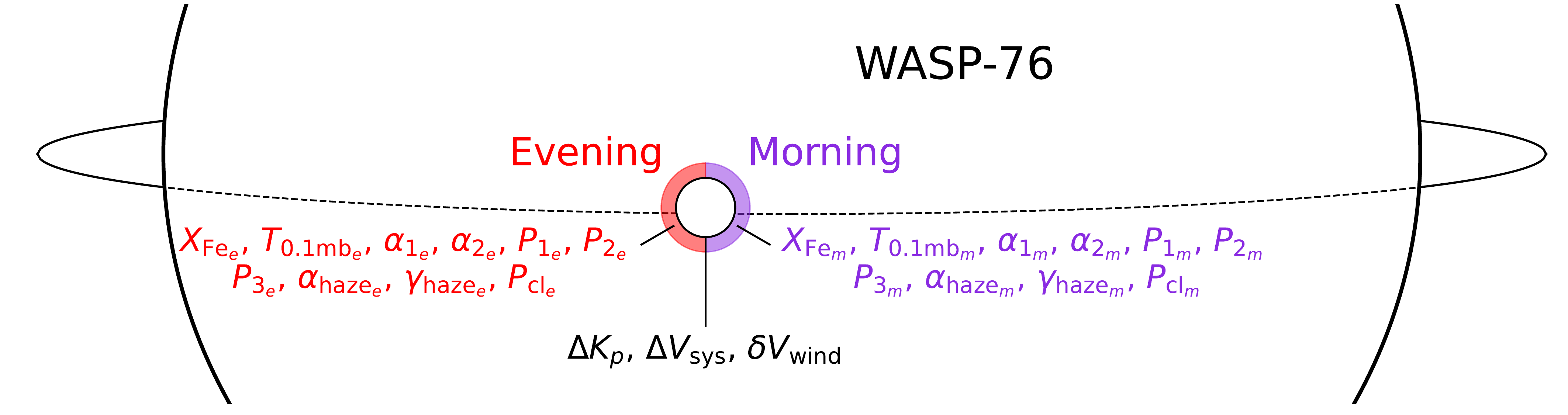}
    \caption{Schematic of the transit of WASP-76~b and the retrieval parameters for HyDRA-2D. For each of the evening and morning limbs we retrieve the gaseous iron abundance with $X_\mathrm{Fe}$, the temperature profile is set through the six parameters $T_\mathrm{mb}$, $\alpha_1$, $\alpha_2$, $P_1$, $P_2$ and $P_3$ using the parametrisation in \citet{madhu2009}, and the opacity and haze are parametrised with $\alpha_\mathrm{haze}$, $\gamma_\mathrm{haze}$ and $P_\mathrm{cl}$. We also retrieve $\Delta K_\mathrm{p}$, $\Delta V_\mathrm{sys}$ and $\delta V_\mathrm{wind}$ to constrain the wind profile of the atmosphere (see section~\ref{sec:methods_winds}).} 
\label{fig:hydra2d_schem}
\end{figure*}

\section{Methods}

In this section we discuss the setup for HyDRA-2D. We retrieve 2 nights of observations of the primary eclipse of WASP-76~b with the ESPRESSO spectrograph on the VLT \citep{ehrenreich2020}. The observations span the optical, where Fe has prominent opacity. As these observations show strong variability in both the strength and position of the Fe signal, we must allow our retrieval to be flexible enough to incorporate such dynamics and variability in a computationally efficient manner. This need for efficiency leads us to a new parametrised approach where 3-dimensional effects are encoded as extra parameters in the description of the physical properties of the atmosphere (e.g., temperature profiles, abundances and opacity decks) and the dynamics of the atmosphere (e.g., velocity shifts and broadening of spectral lines). We firstly discuss the line-by-line opacity calculations for Fe used to compute the high resolution model spectra. We then discuss how we separate the terminator into two distinct regions and the retrieval parameters for each side. Following this we discuss how we incorporate the day-night wind and limb darkening, and then describe our analysis of the data and the model sequence.

\subsection{Fe opacity}
We include Fe opacity through computation of the cross section on a grid of the relevant pressures and temperatures using the Kurucz line list \citep{kurucz1995}. This line list provides the $\log(gf)$ value for each transition, where $g$ is the degeneracy and $f$ refers to the oscillator strength. For our work, we firstly compute the line strength in cm$^{-1}$/(atom cm$^{-2}$) for each temperature $T$ on our grid, through
\begin{align}
    S(T) = \frac{gf \pi e^2}{m_ec^2} \frac{\mathrm{exp}(-c_2E_\mathrm{lower}/T)}{Q(T)} (1-\mathrm{exp}(-c_2\nu_0/T)),
\end{align}
where $\nu_0 = E_\mathrm{upper} - E_\mathrm{lower}$, with the upper and lower state energies of the transition $E_\mathrm{upper}$ and $E_\mathrm{lower}$ given in cm$^{-1}$. The charge of an electron, $e$, the speed of light $c$, and the mass of the electron $m_e$ are all given in c.g.s units, and $c_2 = 1.4387769$~cm~K. In addition, the partition function $Q$ is defined as
\begin{align}
Q(T) &= \sum_j g_j \mathrm{exp}(-c_2E_j/T)
\end{align}
for a state $j$ with degeneracy $g_j$.

We apply line broadening for each of the transitions, which arises from the radiative damping constant (natural broadening) and van der Waals broadening. These result in a Lorentzian line profile at a given frequency $\nu$ (in cm$^{-1}$),
\begin{align}
f_L(\nu-\nu_0) &= \frac{1}{\pi}\frac{\gamma_L}{(\nu-\nu_0)^2+\gamma_L^2}.
\end{align}
Here, $\gamma_L$ is defined as \citep{sharp2007}
\begin{align}
    \gamma_L = \frac{1}{4\pi c}\bigg(\gamma_W N \bigg(\frac{T}{10,000}\bigg)^{0.3} + \gamma_N\bigg),
\end{align}
where $\gamma_W$ and $\gamma_N$ are the van der Waals and natural broadening coefficients respectively and $N$ is the number density of gas, which in our case consists of atomic hydrogen. For our purposes we ignore the Stark broadening of the line given that the atmosphere is not expected to be highly ionised.

In addition to Lorentzian broadening, thermal broadening arising from the temperature of the gas will also contribute to the overall profile. This contribution results in a Gaussian line profile \citep[e.g.,][]{hill2013},
\begin{align}
f_G(\nu-\nu_0) &= \frac{1}{\gamma_G \sqrt[]{\pi}}\mathrm{exp}\left(-\frac{(\nu-\nu_0)^2}{\gamma_G^2}\right),\\
\gamma_G &= \sqrt[]{\frac{2k_bT}{m}}\frac{\nu_0}{c},
\end{align}
where $k_b$ is the Boltzmann constant and $m$ is the mass of an Fe atom. The overall line shape is thus a convolution between the Lorentzian and Gaussian profile, given by the Voigt profile \citep[e.g.,][]{gandhi2017},
\begin{align}
f_V(\nu-\nu_0,\gamma_L,\gamma_G) &= \int_{-\infty}^{\infty}f_G(\nu'-\nu_0)f_L(\nu-\nu')d\nu'.
\end{align}
Thus the cross section as a function of frequency for each transition line is then computed as
\begin{align}
\sigma_\mathrm{Fe} = S(T)f_V(\nu- \nu_0,\gamma_L,\gamma_G).
\end{align}
The sum of the contributions from each line in the line list gives the total atomic cross section of Fe. For each pressure and temperature on our grid, we compute the cross section at 0.01~cm$^{-1}$ spacing, corresponding to a resolution of $R=2.5\times10^6$ at 0.4~$\mu$m. Further details on line broadening and computation of the cross section can be found in \citet{gandhi2020_cs}.

We compute the overall opacity arising from Fe through the volume mixing ratio, $X_\mathrm{Fe}$, which we leave as a free parameter as listed in Table~\ref{tab:priors}. The extinction coefficient $\chi$ (in m$^{-1}$) is given by
\begin{align}
    \chi(P,T,\lambda) = X_\mathrm{Fe} N \sigma_\mathrm{Fe}(P,T,\lambda),
\end{align}
where $\sigma_\mathrm{Fe}$ is the total cross section of Fe and $N$ is the total number density of gas. The contribution to the optical depth $\mathrm{d}\tau$ arising from an element of the atmosphere of length $\mathrm{d}z$ is then
\begin{align}
    \mathrm{d}\tau = \chi(P,T, \lambda) \mathrm{d}z.
\end{align}
Note that we include two free parameters for the Fe abundance, one for the morning and one for the evening side, as discussed below.

\begin{table}
    \centering
    \begin{tabular}{c|c|c}
\textbf{Parameter}              & \multicolumn{2}{c}{\textbf{Prior Range}}\\
 & {Evening side} & {Morning side} \\
\hline
$\log(X_\mathrm{Fe})$ & -15 $\rightarrow$ -2 & -15 $\rightarrow$ -2\\
$T_\mathrm{0.1mb}$ / K & 1500 $\rightarrow$ 3800 & 1500 $\rightarrow$ 3800\\
$\alpha_1\, /\, \mathrm{K}^{-\frac{1}{2}}$ & 0 $\rightarrow$ 1& 0 $\rightarrow$ 1\\
$\alpha_2\, /\, \mathrm{K}^{-\frac{1}{2}}$ & 0 $\rightarrow$ 1& 0 $\rightarrow$ 1\\
$\log(P_1 / \mathrm{bar})$ & -7 $\rightarrow$ 2& -7 $\rightarrow$ 2\\
$\log(P_2 / \mathrm{bar})$ & -7 $\rightarrow$ 2& -7 $\rightarrow$ 2\\
$\log(P_3 / \mathrm{bar})$ & -2 $\rightarrow$ 2& -2 $\rightarrow$ 2\\
$\log(\alpha_\mathrm{haze})$ & -4 $\rightarrow$ 6& -4 $\rightarrow$ 6\\
$\gamma_\mathrm{haze}$ & -20 $\rightarrow$ -1& -20 $\rightarrow$ -1\\
$\log(P_\mathrm{cl}/\mathrm{bar})$ & -7 $\rightarrow$ 2& -7 $\rightarrow$ 2\\
\hline
$\Delta K_\mathrm{p}$ / kms$^{-1}$ &\multicolumn{2}{c}{-20 $\rightarrow$ 20} \\
$\Delta V_\mathrm{sys}$ / kms$^{-1}$ &\multicolumn{2}{c}{-20 $\rightarrow$ 20} \\
$\delta V_\mathrm{wind}$ / kms$^{-1}$ &\multicolumn{2}{c}{1 $\rightarrow$ 20}\\
    \end{tabular}
    \caption{Parameters and uniform prior ranges for HyDRA-2D retrieval. We retrieve separate Fe abundances, temperature profile, and opacity/haze decks for each of the morning and evening sides of the terminator. The $\Delta K_\mathrm{p}$, $\Delta V_\mathrm{sys}$ and $\delta_\mathrm{wind}$ parameters are the same for both sides (see Figure~\ref{fig:hydra2d_schem}).}
    \label{tab:priors}
\end{table}

\subsection{Separating the morning and evening limb}
For our work we assume separate physical properties for the leading limb (or morning side) and trailing limb (or evening side) of the terminator (see Figure~\ref{fig:hydra2d_schem}). This is because we are able to separate the spectrally contributing regions of the terminator due to their difference in velocity space given the high resolution (R=140,000) and signal-to-noise. We therefore retrieve separate Fe volume mixing ratios for the morning and evening sides of the atmosphere. In addition, we also retrieve separate pressure-temperature profiles and opacity decks/hazes for each half of the terminator.

\begin{figure*}
\centering
	\includegraphics[width=\textwidth,trim={0cm 0cm 0cm 0},clip]{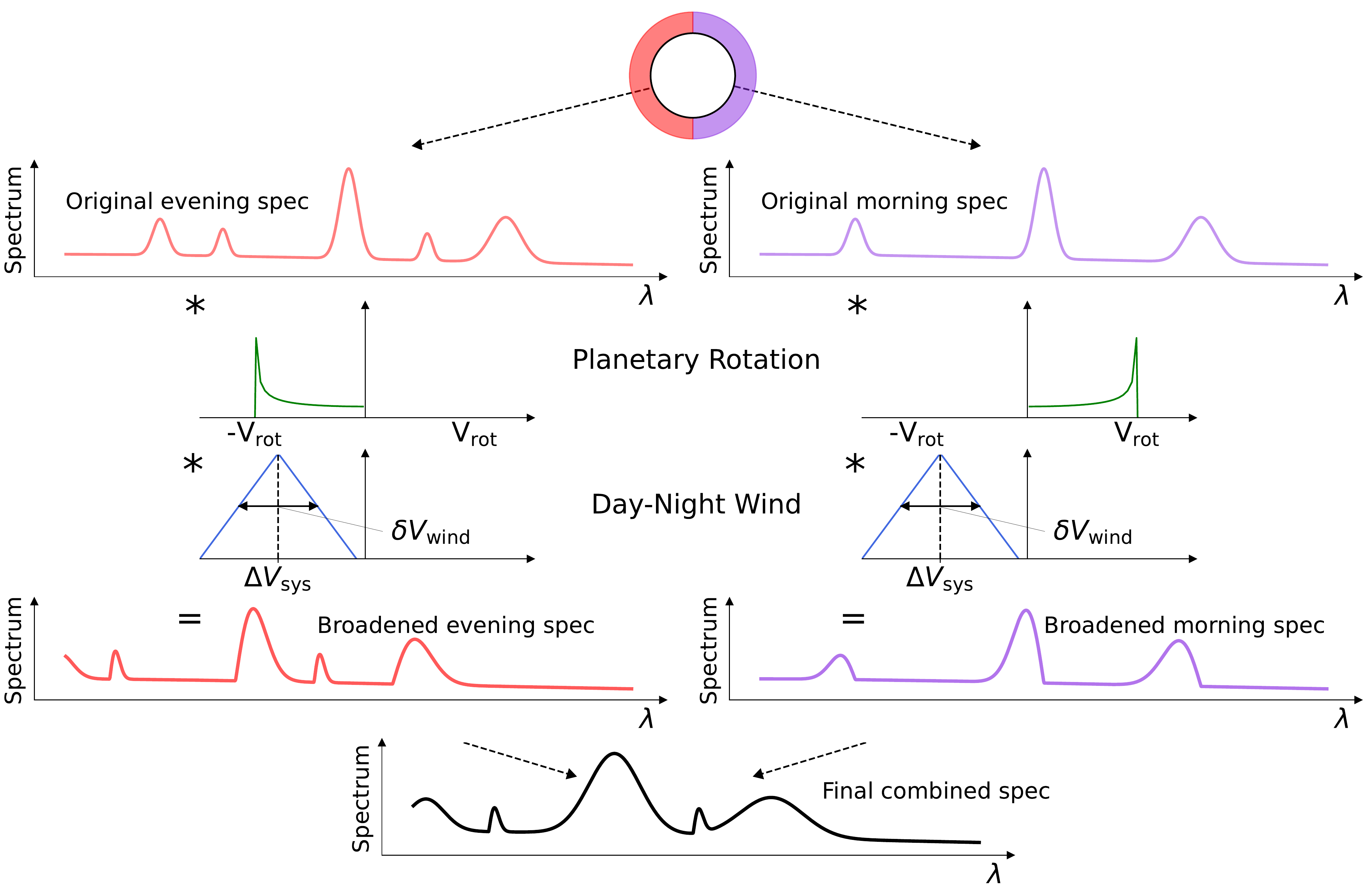}
    \caption{Schematic of the broadening prescription used for HyDRA-2D. We firstly compute the unbroadened evening and morning transit depths using the parameters for each side (see Table~\ref{tab:priors}). We then convolve this spectrum with the rotational broadening kernel for each side. The evening side is rotating towards the observer and the morning side is rotating away, and thus are blue- and red-shifted respectively. We then shift and broaden both sides by a day-night wind with $\Delta V_\mathrm{sys}$ and $\delta V_\mathrm{wind}$, which are free parameters in the retrieval. We then combine the spectra for each limb after incorporating the stellar limb darkening and model reprocessing, as discussed in sections~\ref{sec:methods_limb_dn} and \ref{sec:data_analysis}.} 
\label{fig:broaden_schem}
\end{figure*}

\subsubsection{Temperature profile}
We parametrise the temperature profile for each side using the method described in \citet{madhu2009}. This requires breaking the atmosphere into 3 distinct regions, with the top two regions capable of having a varying temperature profile with pressure, and the final deepest layer assuming an isothermal temperature expected to be common at high pressures on hot and ultra-hot Jupiters \citep[e.g.,][]{fortney2008}. We thus obtain the following parametrisation for $T$ as a function of pressure $P$,
\begin{align}
    T(P) = \begin{cases}
    T_0 + \Big(\frac{\ln{P/P_0}}{\alpha_1}\Big)^2 & \text{if $P\leq P_1$},\\
    T_2 + \Big(\frac{\ln{P/P_2}}{\alpha_2}\Big)^2 & \text{if $P_1 < P \leq P_3$},\\
    T_3 & \text{if $P>P_3$},
    \end{cases}
\end{align}
and where we enforce $P_1 \leq P_3$. We additionally fix $P_0 = 10^{-7}$~bar as the top of the model atmosphere. At lower pressures, the assumptions of hydrostatic equilibrium and local thermodynamic equilibrium begin to break down. We allow the value of $P_2$ to vary between $P_0 \leq P_2 \leq P_3$. Importantly, such a constraint allows for a thermal inversion in layer 2 in addition to isothermal and non-inverted temperature profiles. This is done to allow the retrieval to determine the best fitting $P-T$ profile without the requirement that the temperature be monotonically decreasing with altitude. By enforcing continuity of the temperature between the 3 layers, we can reduce this set of equations to just 6 free parameters, a temperature value at any given pressure, $\alpha_1$, $\alpha_2$, $P_1$, $P_2$ and $P_3$. For our retrieval we retrieve the temperature at 0.1~mbar, $T_\mathrm{0.1mb}$, which is representative of a typical photospheric pressure in transmission geometry for such high resolution observations. We have also verified our results by using the temperature at the top of the model atmosphere as our free parameter, as well as with the 0.1~bar temperature as a free parameter, to ensure this did not impact our constraints. The prior ranges for each parameter is shown in Table~\ref{tab:priors}. We use different sets of parameters for the evening side and morning sides, which ensures that the temperature constraints for one side are independently constrained of the other, and hence overall we have 12 free parameters in the retrieval. Further details on the parametrisation of the $P-T$ profile can be found in \citet{madhu2009}.

\subsubsection{Grey opacity deck and haze}
In addition to the opacity resulting from the presence of gaseous Fe, other sources of opacity may play a key role in determining the spectrum. These opacities arise from of a number of factors, such as other molecular and atomic species in the atmosphere, and non-Rayleigh scattering from small particles/hazes in the atmosphere. These result in two distinct sources of extinction in the atmosphere, namely that of a wavelength-independent opacity deck and a wavelength-dependant super-Rayleigh haze. For our prescription we adopt a one sector approach with homogeneous opacity and haze parameters \citep[e.g.,][]{macdonald2017, welbanks2021} for each of the morning and evening limbs. We add this source of opacity into the overall extinction coefficient of the atmosphere, $\chi$, as
\begin{align}
    \chi(P,\lambda) = \begin{cases}
    \alpha_\mathrm{haze} N\sigma_0\big(\frac{\lambda}{\lambda_0}\big)^{\gamma_\mathrm{haze}} & \text{if $P\leq P_\mathrm{cl}$},\\
    \infty & \text{otherwise},
    \end{cases}
\end{align}
where $N$ is the total number density of gas, $\sigma_0 = 5.31 \times 10^{-31}$~m$^2$ and $\lambda_0 = 0.35$~$\mu$m. Hence, the free parameters are $\alpha_\mathrm{haze}$, $\gamma_\mathrm{haze}$ and $P_\mathrm{cl}$, whose prior ranges are given in Table~\ref{tab:priors}. This therefore results in 6 free parameters in our retrieval given that we assume the two sides of the atmosphere are independent. The overall opacity of each side of the atmosphere is thus the sum of the contributions arising from Fe line absorption and the opacity deck/haze. Further details on the opacity deck/haze prescription can be found in \citet{welbanks2021}.

\subsubsection{Computing the unbroadened spectrum}
We compute the unbroadened spectrum using numerical radiative transfer for each side of the atmosphere \citep[e.g.,][]{welbanks2021}. This involves computation of the transit depth $\Delta_e$ and $\Delta_m$ for the evening and morning sides, and is given by
\begin{align}
    \Delta_e(\lambda) = \frac{1}{2}\, \frac{1}{R_\mathrm{s}^2}\Bigg(R_\mathrm{p}^2 + 2 \int_{R_\mathrm{p}}^{\infty}b_e & (1-e^{-\tau_e(\lambda)}) \, \mathrm{d}b_e &\nonumber \\
    & - 2\int_{0}^{R_\mathrm{p}}b_e e^{-\tau_e(\lambda)} \, \mathrm{d}b_e\Bigg),\\
    \Delta_m(\lambda) = \frac{1}{2}\, \frac{1}{R_\mathrm{s}^2}\Bigg(R_\mathrm{p}^2 + 2 \int_{R_\mathrm{p}}^{\infty}b_m & (1-e^{-\tau_m(\lambda)}) \, \mathrm{d}b_m &\nonumber \\
    & - 2\int_{0}^{R_\mathrm{p}}b_m e^{-\tau_m(\lambda)} \, \mathrm{d}b_m\Bigg),
\end{align}
where $R_\mathrm{p}$ and $R_\mathrm{s}$ refer to the radius of the planet and star respectively, $\tau$ refers to the optical depth, and the impact parameter is given by $b$. The subscripts $e$ and $m$ refer to the evening and morning sides, and the factor of 1/2 accounts for the fact that each side is only integrated for half of the limb. The physical properties of each side such as temperature and Fe abundance are variable, and thus the overall optical depth for each side is different. We set our reference radius $R_\mathrm{p} = 1.854 \,R_\mathrm{J}$ at 10$^{-2}$~bar reference pressure, but the chosen pressure level has no effect on the retrieval given the high-pass filtering in the data analysis removes any offset between different chosen pressures (see section~\ref{sec:data_analysis}). The spectrum for each side is then broadened and combined as discussed below.

\begin{figure}
\centering
	\includegraphics[width=0.49\textwidth,trim={0.0cm 0cm 0cm 0},clip]{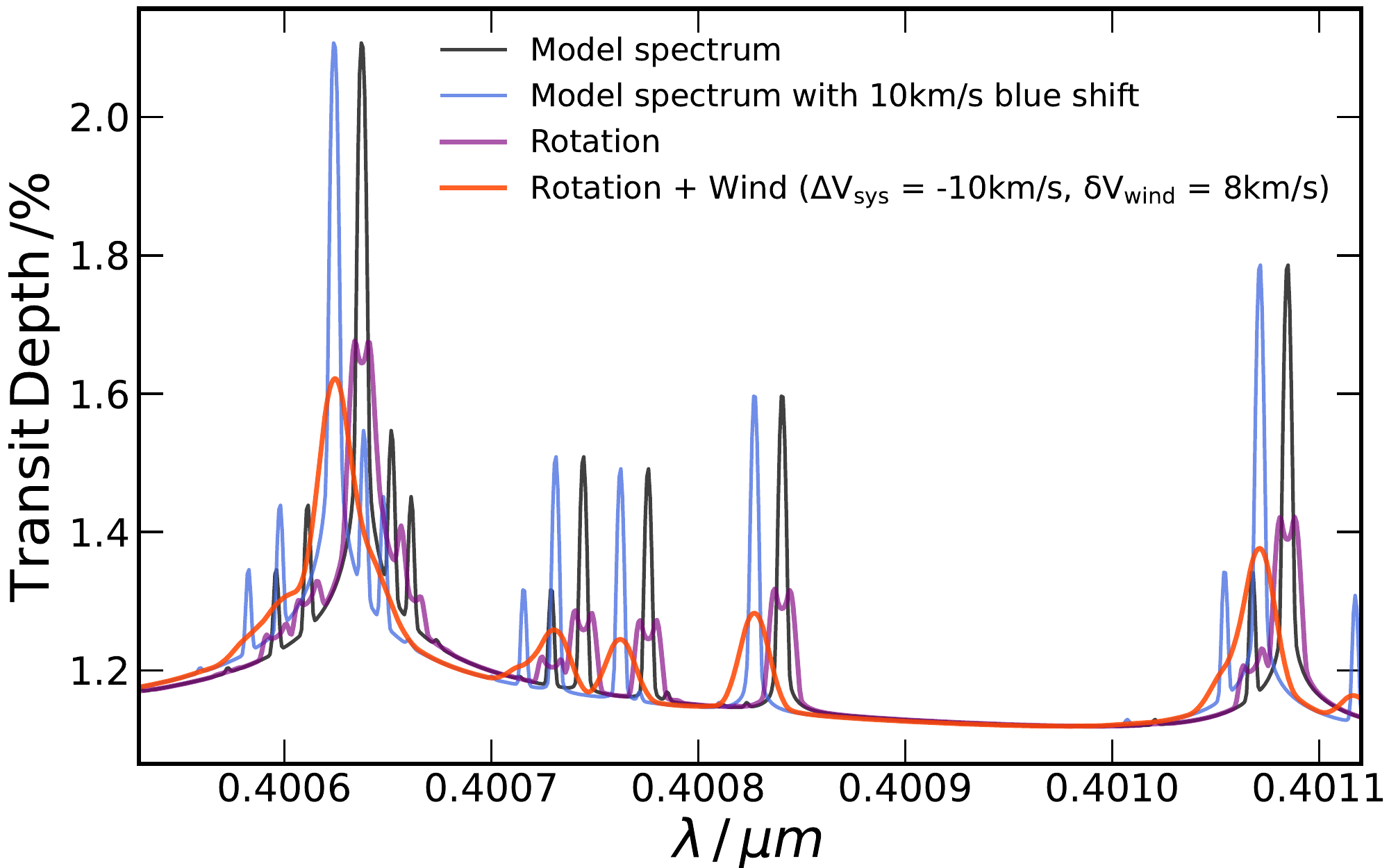}
    \caption{Example spectra in the $\sim$0.4~$\mu$m range showing the effect of rotational and wind broadening on our model. The black line shows the unbroadened spectrum, which we set to be the same for both the evening and morning sides for this illustration. We apply the rotational broadening as shown in purple, followed by a day-night wind with $\Delta V_\mathrm{sys} = -10$~km/s and $\delta V_\mathrm{wind} = 8$~km/s as shown in red. We also show an unbroadened spectrum shifted by -10~km/s in blue.}
\label{fig:spec_br}
\end{figure}

\subsection{Incorporating winds}\label{sec:methods_winds}
Once we have computed the transit spectra for each side of the atmosphere, we must broaden them according to the rotation of the planet and the winds present in the atmosphere. The procedure for this broadening is shown in Figure~\ref{fig:broaden_schem}. This setup was motivated through recent work on general circulation models by \citet{wardenier2021}, which showed that the broadened primary eclipse spectrum is significantly altered and therefore requires careful consideration of both aspects for accurate spectral modelling.

\subsubsection{Planetary rotation}

We firstly begin by applying the planetary rotation to each side. The evening side, or trailing limb, is rotating towards the observer and the morning side, or leading limb, is rotating away. Therefore, the evening and morning sides will be blue- and red-shifted respectively according to the rotation of the planet. However, the polar regions of the terminator will be rotating more slowly than the equator and hence there is also a contribution arising at lower velocities. For a rotation velocity, $V_\mathrm{rot}$, the full rotation kernel is given by
\begin{align}
    R(x) = 
    \begin{cases}
    \frac{1}{\pi \sqrt{1 - x^2/V_\mathrm{rot}^2}} & \text{if $|x|<V_\mathrm{rot}$}\\
    0 & \text{otherwise}.
    \end{cases}
\end{align}
We only apply half of the kernel for each half of the terminator, as shown in Figure~\ref{fig:broaden_schem}. This convolves our unbroadened spectrum according to the respective kernel for each side, which also results in a net shift for each side. We keep the rigid-body rotation kernel fixed for each model in our retrieval as the rotational velocity of WASP-76~b is well known given that the planet is expected to be tidally locked \citep[e.g.,][]{showman2002}.

An example spectrum highlighting the effect of the broadening is shown in Figure~\ref{fig:spec_br} for a small region near $\sim$0.4~$\mu$m. The rotation results in a split of the overall line profiles in the spectrum, with two distinguishable peaks in the rotationally broadened spectrum arising from contributions from each half of the terminator. In addition, the spectrum is more spread in wavelength and therefore the sharp peak arising from Fe line opacity in the unbroadened spectrum has reduced in extent. Such rotation is therefore vital to account for as it significantly alters the line profile, and may otherwise lead to biased estimates of abundance. 

\subsubsection{Day-night wind}

In addition to the rotation, a day-night wind also contributes to the broadening of the overall spectrum. General circulation models have shown that the highest velocity winds are at the lowest pressures, with deeper regions having almost no wind present \citep[e.g.,][]{kempton2012, wardenier2021}. Therefore incident rays of light from the host star that pass through the atmosphere first encounter a strong wind at the top of the atmosphere, then pass through to higher pressures with a lower wind speed, and finally exit out of the atmosphere through the lower pressures with higher wind speeds again. This path thus results in a net blue-shift as well as a spread in the spectrum due to the range in velocities encountered by the rays of light passing through the atmosphere. In addition, there will also be a spread caused by the wind speed even if it is constant with depth, given that the projected velocity into the observer's reference frame varies as the path travels through the atmosphere. There may also be an additional contribution arising from variation of the winds with latitude. 

To account for these physical mechanisms and processes, we parametrise the wind with a velocity shift from the known systemic velocity, $\Delta V_\mathrm{sys}$, and a spread in its value given by the full-width half-maximum (FWHM) $\delta V_\mathrm{wind}$ (see Figure~\ref{fig:broaden_schem}). The $\Delta V_\mathrm{sys}$ parameter is often included in high-resolution analyses \citep[e.g.,][]{birkby2013, pino2020, giacobbe2021} given that the systemic velocity often has some measurement uncertainty which we may be sensitive to. However, in our case, the $V_\mathrm{sys}$ parameter is very precisely known at -1.17$\pm$0.02 km/s \citep{ehrenreich2020} and thus we can assume that any deviation $\Delta V_\mathrm{sys}$ can be attributed to the winds in the atmosphere. We also include one additional parameter, the deviation from the planet's known orbital velocity, $\Delta K_\mathrm{p}$, as this often has a degeneracy with the $\Delta V_\mathrm{sys}$ value. We therefore have three additional parameters in HyDRA-2D to account for the wind and its uncertainty, $\Delta K_\mathrm{p}$, $\Delta V_\mathrm{sys}$ and $\delta V_\mathrm{wind}$ as shown in Table~\ref{tab:priors}.

Figure~\ref{fig:spec_br} shows the effect of applying the wind broadening. We now see a net blue-shift in the spectrum as a result of the wind $\Delta V_\mathrm{sys}$, and an additional spread in the line profile as a result of $\delta V_\mathrm{wind}$. Note that the spectrum is significantly more spread than the unbroadened spectrum shifted by 10~km/s, and therefore the wind and rotation have a strong effect on our retrieved results. The resulting spectral line profile for this broadened model with rotation and wind is similar to the type of line distortions and shifts predicted by GCMs \citep[e.g.,][]{showman2013, rauscher2014}. We are therefore confident that our parametrisation can address the diversity of potential line profiles.

\begin{figure}
\centering
	\includegraphics[width=0.35\textwidth,trim={0.0cm 0cm 0cm 0},clip]{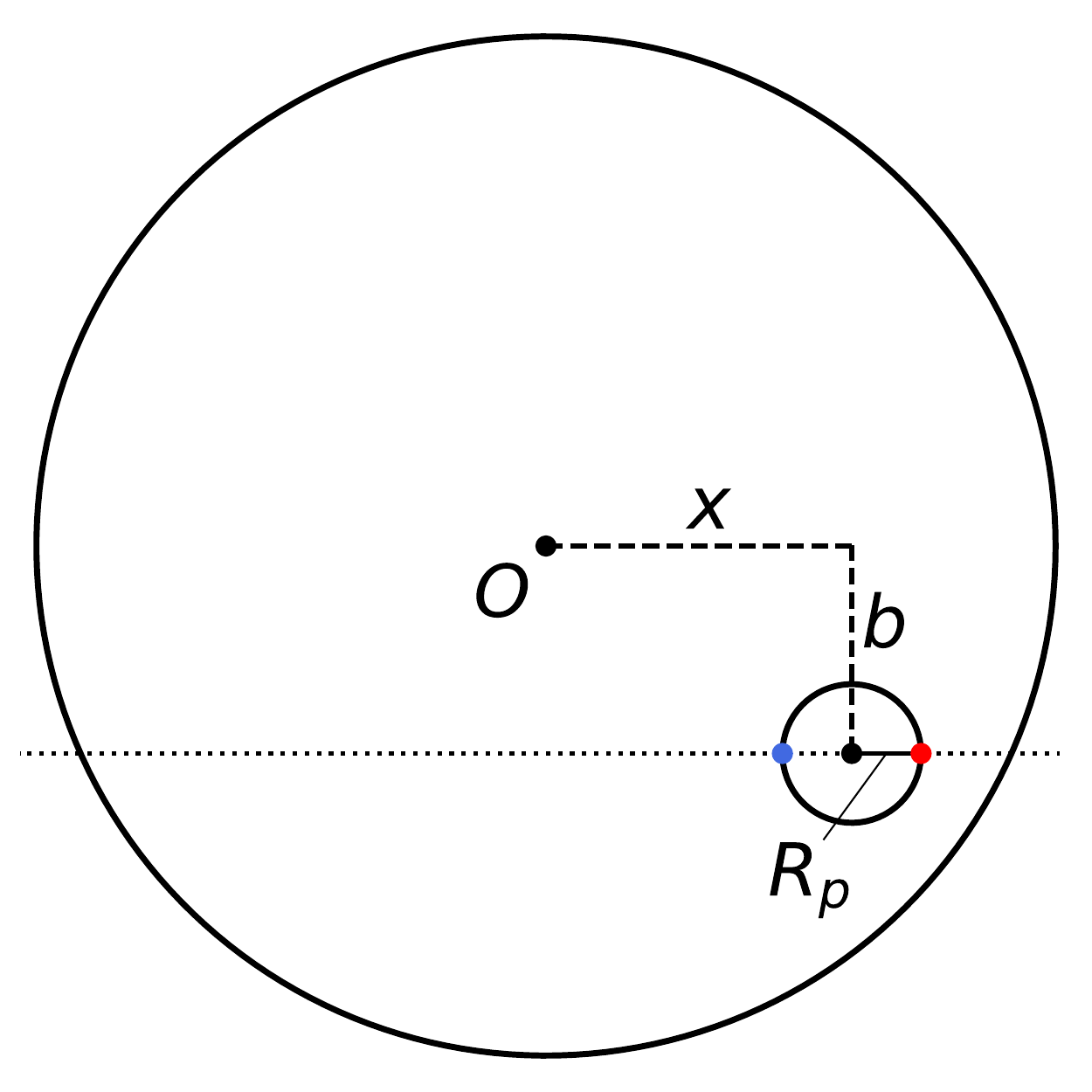}
    \caption{Schematic showing the setup for the limb darkening calculations. The outer circle shows the star and the inner circle showing the planet with radius $R_\mathrm{p}$, with its path during the transit given by the dotted line. The impact parameter of the transit is given by $b$. For the evening and morning side limb darkening factors we calculate the brightness at the blue and red points respectively (see section~\ref{sec:methods_limb_dn}).}
\label{fig:limb_dn_schem}
\end{figure}

\subsection{Stellar limb darkening}\label{sec:methods_limb_dn}
For each spectrally broadened half of the terminator, we must also apply limb darkening, given that the brightness of WASP-76's stellar surface varies significantly in the optical. As the retrievals are over the first and last quarter of the transit, the brightness and therefore the relative contributions of the morning and evening side will thus vary with the orbital phase. 

Figure~\ref{fig:limb_dn_schem} shows the setup to compute the limb darkening for each side of the atmosphere. We first compute the value of $x$ for a given phase of the observations $\phi$, normalised by the stellar radius $R_\mathrm{s}$, given by \begin{align}
    x = \frac{a \sin(2\pi\phi)}{R_\mathrm{s}},
\end{align}
where $a$ refers to the semi-major axis of the orbit. We must then add and subtract $R_\mathrm{p}/R_\mathrm{s}$ for the morning and evening sides respectively. For simplicity we do not integrate the contribution of the limb darkening over the half-annulus of each side, but take its value at the red and blue points in Figure~\ref{fig:limb_dn_schem} for the morning and evening. This is justified given that there is no significant variability in the limb over the half-annuli and given that the modal contribution to the spectrum comes from these points. For each side we thus obtain
\begin{align}
    x_\mathrm{e} = \frac{a\sin(2\pi\phi)}{R_\mathrm{s}} - \frac{R_\mathrm{p}}{R_\mathrm{s}},\\
    x_\mathrm{m} = \frac{a\sin(2\pi\phi)}{R_\mathrm{s}} + \frac{R_\mathrm{p}}{R_\mathrm{s}},
\end{align}
with the morning side given by $x_\mathrm{m}$ and the evening side given by $x_\mathrm{e}$. We then compute the radial distance $\mu$ of the points from the centre of the star,
\begin{align}
    \mu_\mathrm{e} = \sqrt{1 - x_\mathrm{e}^2 - b^2},\\
    \mu_\mathrm{m} = \sqrt{1 - x_\mathrm{m}^2 - b^2}.
\end{align}
The impact parameter $b$ is assumed to be 0.027 \citep{ehrenreich2020}. We then compute the overall limb darkening function given the quadratic limb darkening coefficients $u_1$ and $u_2$,
\begin{align}
    I_\mathrm{e}(\phi) = \begin{cases}
    1 - u_1(1-\mu_\mathrm{e}) - u_2(1-\mu_\mathrm{e})^2 & \text{if $x_\mathrm{e}^2 \leq 1 - b^2$},\\
    0 & \text{elsewhere}.
    \end{cases}
\end{align}
Similarly for the morning side we obtain
\begin{align}
    I_\mathrm{m}(\phi) = \begin{cases}
    1 - u_1(1-\mu_\mathrm{m}) - u_2(1-\mu_\mathrm{m})^2 & \text{if $x_\mathrm{m}^2 \leq 1 - b^2$},\\
    0 & \text{elsewhere}.
    \end{cases}
\end{align}
For our work we have taken $u_1 = 0.393$ and $u_2 = 0.219$ \citep{ehrenreich2020}. 

We obtain the overall spectrum by simply scaling the model for each side by the value of $I$ and summing (see equation~\ref{eqn:final_spec} below). Hence, with this limb-darkening model we are able to account for the variation in the contributions for each side across the transit; the evening side is the dominant contributor during the latter half of the transit and vice versa for the morning in the first half. In addition, the model ensures that at ingress and egress we obtain a contribution from just one side in the overall spectrum.

\subsection{Data analysis and model reprocessing}\label{sec:data_analysis}

\begin{figure*}
\centering
	\includegraphics[width=0.9\textwidth,trim={0cm 0cm 0cm 0},clip]{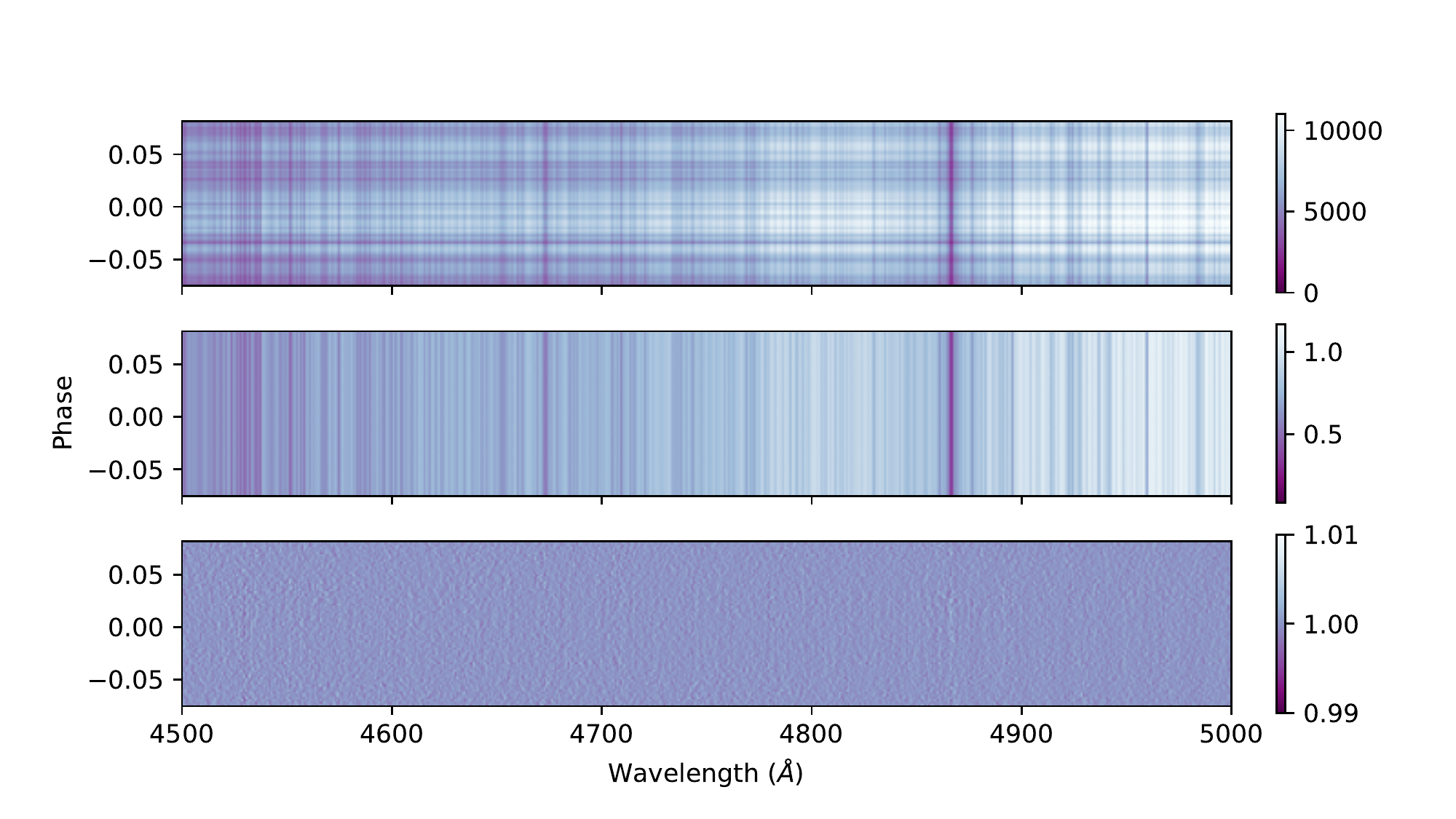}
    \caption{Example showing the data analysis steps that were performed. The top panel shows a small slice in wavelength of the pipeline-reduced ESPRESSO spectra taken during the transit on 2018-10-30. Each of the 69 rows is a different spectrum that has been shifted to the star's rest frame. Vertical dark lines are stellar absorption lines. The colour bar shows the pixel values of the spectra. In the middle panel, the spectra have been corrected for changes in signal-to-noise ratio and any difference in the blaze function that occur throughout the night, which removes the horizontal banding seen in the top panel. The bottom panel shows the residuals that are left after the stellar spectrum was removed by dividing by the average out-of-transit spectrum. The exoplanet's signal is hidden in the noise, and is extracted using cross correlation.}
\label{fig:data_analysis}
\end{figure*}

As the data are the same that were used in \citet{kesseli2022}, the data are cleaned and processed in a similar way. We summarise the steps here (see Figure \ref{fig:data_analysis}), but for more details on the observing strategy and initial data reduction see \citet{ehrenreich2020}, and for more details on the spectral cleaning steps see \citet{kesseli2022}. 

We downloaded the pipeline reduced ESPRESSO spectra from the Data and Analysis Center for Exoplanets (DACE) database\footnote{\url{https://dace.unige.ch/dashboard/index.html}}. Before any other steps, we correct the spectra for telluric absorption lines due to H$_2$O and O$_2$ using molecfit \citep{smette2015}. Molecfit performs especially well for shallow absorption lines, but is known to return poor results for the line cores of the strongest absorption lines \citep[e.g.,][]{allart2017}. For this reason, we mask all pixels where the transmission through the atmosphere is less than 40\%. We then correct for velocity shifts due to the reflex motion of the star ($K_*$), the barycentric velocity ($v_{bar}$), and the system velocity ($v_{sys}$). Next, we perform a 5-$\sigma$ clipping to remove any spurious flux spikes due to cosmic rays (top panel of Figure \ref{fig:data_analysis}). To remove any broadband noise present in the spectra while preserving the overall shape, we follow \citet{merritt2020} and place the spectra on a ``common" blaze using a high-pass filter with a width of 200 pixels ($\sim$100 km s$^{-1}$; middle panel of Figure \ref{fig:data_analysis}). Finally, we interpolate each spectrum onto the same wavelength grid and divide each spectrum by the average out-of-transit stellar spectrum to remove the contribution from host star (bottom panel of Figure \ref{fig:data_analysis}). 

To process our model, we must reproduce the model data cube, beginning first by Doppler shifting the spectrum at each phase by its velocity, $V_\mathrm{p}$. At a given phase $\phi(t)$ the shift is given by
\begin{align}
    V_\mathrm{p}(\phi) = \Delta V_\mathrm{sys} + (K_\mathrm{p} + \Delta K_\mathrm{p})\sin(2\pi\phi(t)),
\end{align}
where $K_\mathrm{p} = 196$~km/s is the planet's rotation velocity. The $\Delta V_\mathrm{sys}$ parameter is the day-night wind velocity and represents an overall shift in the planet's spectrum. We also include $\Delta K_\mathrm{p}$ to account for uncertainties in the measured value of $K_\mathrm{p}$ and because there is often a degeneracy between the $K_\mathrm{p}$ and $V_\mathrm{sys}$ parameters as a result of the limited phase range that we are exploring during the transit. The new wavelengths of the shifted spectrum are
\begin{align}
    \lambda(\phi) &= \lambda_0 \sqrt{ \frac{1+V_\mathrm{p}(\phi)/c}{1-V_\mathrm{p}(\phi)/c}},
\end{align}
where $\lambda_0$ is the wavelength of the unshifted spectrum. Therefore, the final combined spectrum for the terminator is given by
\begin{align}
    \Delta_\mathrm{shift}(\phi) = \Delta_e^{'} (\lambda(\phi)) I_e(\phi) + \Delta_m^{'} (\lambda(\phi)) I_m(\phi),\label{eqn:final_spec}
\end{align}
where $I_e$ and $I_m$ refer to the stellar limb darkening factors, and $\Delta_e^{'}$ and $\Delta_m^{'}$ refer to the evening and morning side transit depths which have been rotationally and wind broadened and shifted to the new wavelengths $\lambda(\phi)$ for each phase. One important final step we take with our model is to apply a high-pass filter, which removes any slowly varying trends in the model and reproduces the data sequence. The high-pass filter also acts to remove the continuum from the spectra, which means that we are only sensitive to the spectral line features which are generated above the continuum.

Once we have a processed spectrum for the atmosphere, we must compare this to the observations of WASP-76~b and determine the likelihood. We use the cross-correlation to log-likelihood (CC-to-logL) map of \citet{brogi2019} to compare our model to the data. We do this by computing
\begin{align}
    \log L = - \frac{N}{2} \log(s_f^2 + s_g^2 - 2R),
\end{align}
where
\begin{align}
    s_f^2 &= \frac{1}{N}\sum_{n=0}^N f^2(n),\\
    s_g^2 &= \frac{1}{N}\sum_{n=0}^N g^2(n-s),\\
    R &= \frac{1}{N} \sum_{n=0}^N f(n)g(n-s).
\end{align}
Here, $f$ and $g$ are the data and model respectively for a model with $N$ spectral points, and $s$ represents the wavelength offset. This form of the likelihood accounts for the overall line shape as well as position; models which poorly match the observations or are scaled incorrectly will be penalised by the $R$ term and thus less favoured. We have two nights of observations for WASP-76~b, and so we sum the likelihood for each night to obtain the overall value. The method in \citet{brogi2019} also prescribes reprocessing each tested model, i.e. reproducing the effects of the data analysis on the model spectrum. This reprocessing is necessary in the infrared because algorithms used to remove telluric lines are known to alter the exoplanet spectrum as well. However, our analysis in the optical is based on the in-transit / out-of-transit approach and does not distort the planetary signal, and therefore the depth and shape of planet's spectral lines are preserved relative to the stellar spectrum. We can therefore compare each model to the observations directly, without significant reprocessing of the model. Further details on the likelihood method and derivation can be found in \citet{brogi2019}. 

\subsection{Retrieval setup}

We retrieve the high-resolution observations of WASP-76~b, with two transits obtained on 02/09/2018 and 30/10/2018 with the ESPRESSO spectrograph on the VLT \citep{ehrenreich2020}. These observations are some of the highest signal-to-noise atmospheric observations with high-resolution spectra, obtained at a spectral resolution of R=140,000 in the optical. We use the 0.4-0.8~$\mu$m wavelength range for our retrievals. We set up HyDRA-2D with 23 free parameters, which are given in Table~\ref{tab:priors}. This includes 10 parameters for each of the morning and evening sides, representing the Fe abundance, temperature profile (6 parameters), and opacity and haze deck (3 parameters), and 3 additional parameters which are consistent for both sides, $\Delta K_\mathrm{p}$, $\Delta V_\mathrm{sys}$ and $\delta V_\mathrm{wind}$. Each spectral model is generated at a spectral resolution of R=500,000, assuming a planet radius of $1.854 \,R_\mathrm{J}$, planet mass of $0.894 \, M_\mathrm{J}$ and stellar radius $1.756\, R_\mathrm{\odot}$. We use the Nested Sampling algorithm MultiNest \citep{feroz2008, feroz2009, buchner2014} for parameter estimation with HyDRA-2D, adopting 1000 live points.

\begin{figure}
\centering
	\includegraphics[width=0.46\textwidth,trim={0.0cm 0cm 0cm 0},clip]{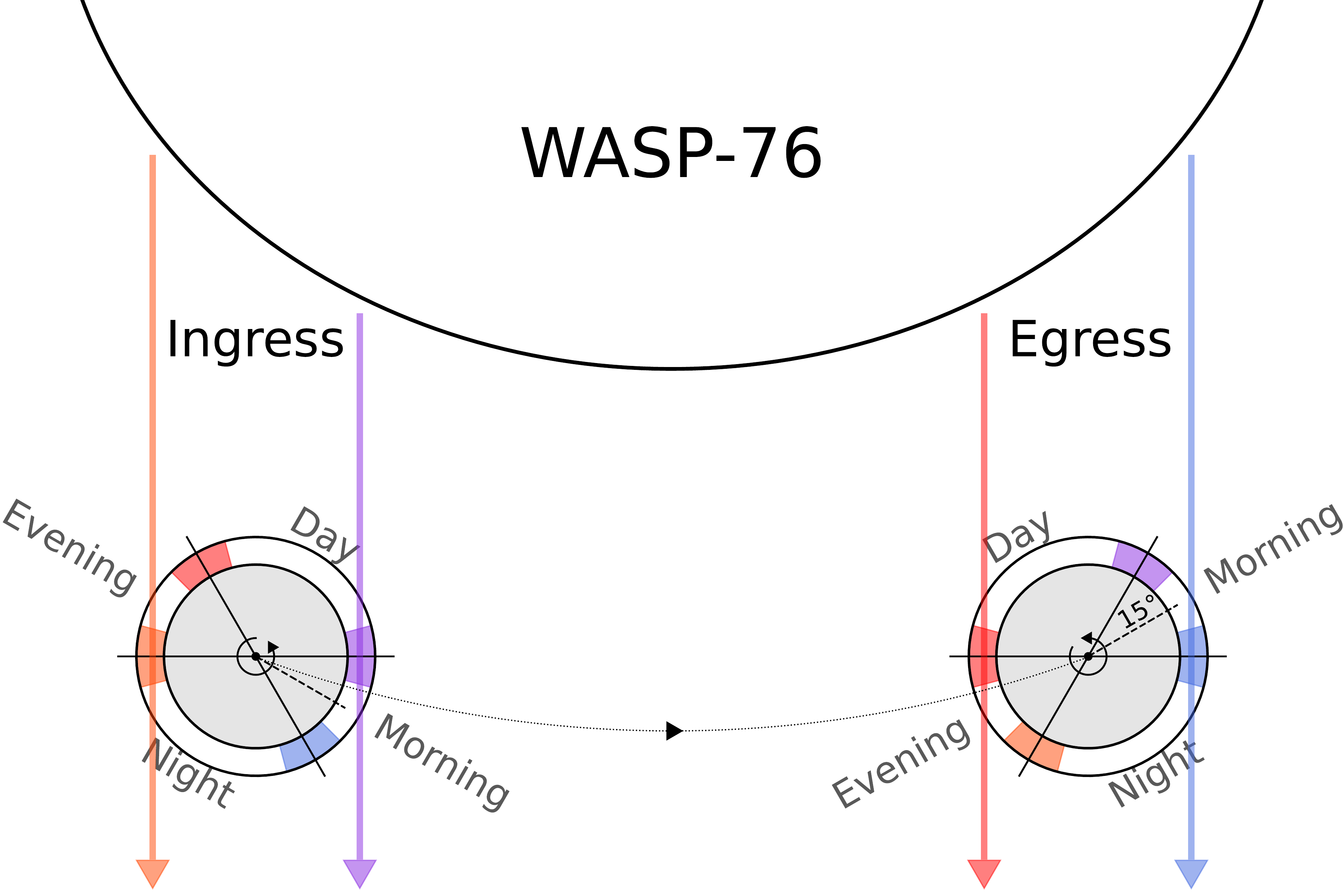}
    \caption{Schematic showing the different regions of the morning (leading) and evening (trailing) limb probed during the transit. During ingress, the stellar light passes through the day side of the morning and the night side of the evening, which then changes as the planet rotates during its transit. During egress the stellar rays pass through more of the say side of the evening and the night side of the morning.}
\label{fig:rot_schematic}
\end{figure}

We perform two separate retrievals, in the $\phi = -0.04 \,\text{-}\, -0.02$ and $\phi = 0.02 \,\text{-}\, 0.04$ phase ranges, leaving out the phases in between due to the Doppler shadow of the star. The transit of the planet occurs between $-0.0432\leq \phi \leq 0.0432$ and the whole planet is in the transit between the phases of $-0.035\leq \phi \leq 0.035$. We choose to use phases up to $\pm0.04$ for our retrieval as at least one side of the planet is in the transit within this range. We have performed tests varying the upper phase range but we did not see any significant difference in the retrieved parameters given that the signal is weaker near ingress and egress due to the dimmer stellar limbs. The two separate retrievals over the phase ranges probe differing regions of the atmosphere due to the planet's rotation angle of $\sim30^\circ$ during the transit (see Figure~\ref{fig:rot_schematic}). This allows us to explore variations within the morning and evening side as the transit progresses and compare any differences. 

\begin{figure*}
\centering
	\includegraphics[width=\textwidth,trim={0cm 0cm 0cm 0},clip]{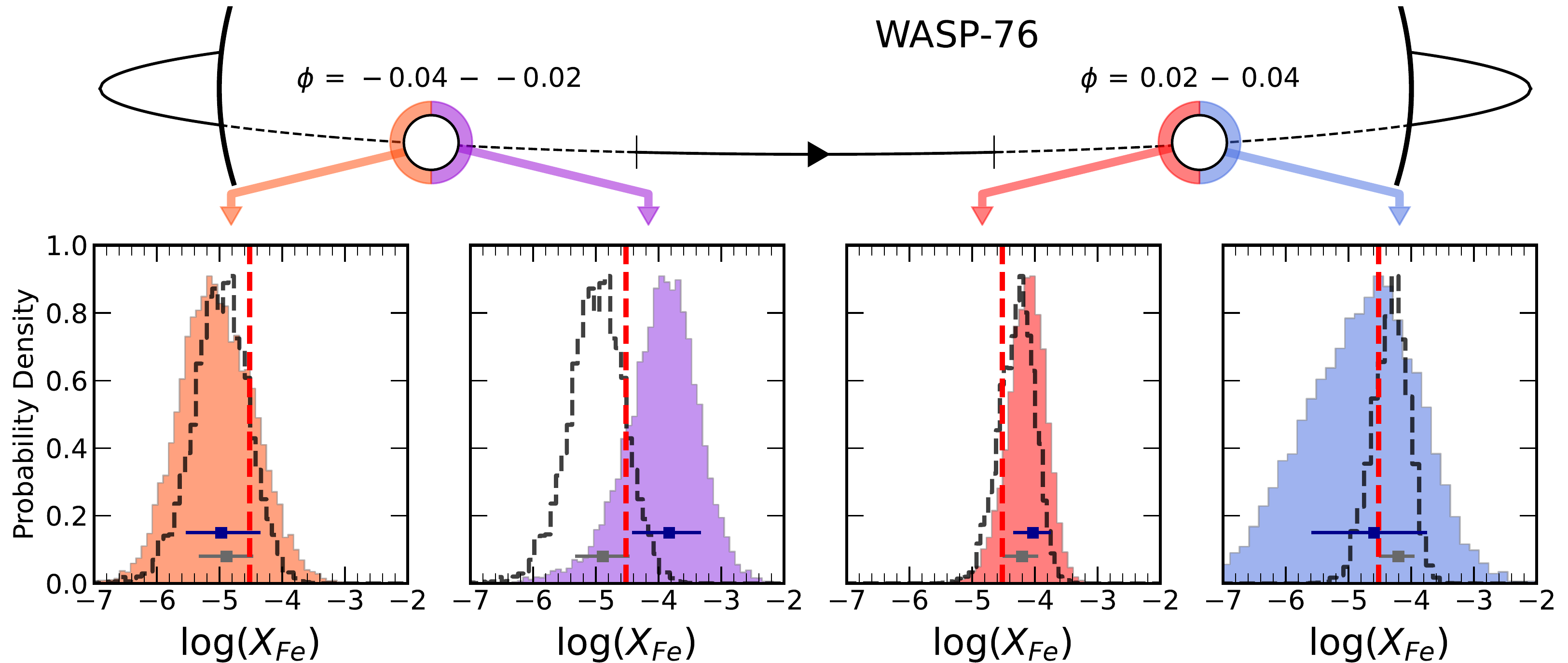}
    \caption{Retrieved Fe constraints from HyDRA-2D for each side of the planet and for each of the retrievals conducted over the $\phi = -0.04 \,\text{-}\, -0.02$ and $\phi = 0.02 \,\text{-}\, 0.04$ ranges. The orange and red histograms give the constraints for the evening, and the purple and blue show the constraints for the morning. The blue bar indicates the median and $\pm1\sigma$ error bar on the retrieved abundances. We also show the retrieved constraints from the separated-phase 1D retrievals which assume a homogeneous terminator with the dashed black histogram, with the grey bar indicating its 1$\sigma$ confidence interval. We additionally plot the solar value of Fe with the red dashed vertical line, with $\log(X_\mathrm{Fe}) = -4.52$ \citep{asplund2009}.} 
\label{fig:fe_constraints}
\end{figure*}

\begin{table*}
    \centering
    \def\arraystretch{1.5}
    \begin{tabular}{c|c|c|c|c}
         & \multicolumn{2}{c}{$\mathbf{\phi = -0.04 \, - \, -0.02}$} & \multicolumn{2}{c}{$\mathbf{\phi = 0.02 \, - \, 0.04}$}\\
         \cline{2-5}
         & Evening & Morning & Evening & Morning\\
         \hline
         \textbf{combined-phase 1D} & \multicolumn{4}{c}{$-4.33^{+0.26}_{-0.27}$}\\
         \hline
         \textbf{separated-phase 1D} & \multicolumn{2}{c}{$-4.89^{+0.43}_{-0.45}$} & \multicolumn{2}{c}{$-4.20^{+0.26}_{-0.31}$}\\
         \cline{2-5}
         {Preference over combined-phase 1D} & \multicolumn{4}{c}{3.8$\sigma$}\\
         \hline
         \textbf{HyDRA-2D} & $-4.97^{+0.63}_{-0.56}$ & $-3.83^{+0.51}_{-0.59}$ & $-4.03^{+0.28}_{-0.31}$ & $-4.59^{+0.85}_{-1.0}$\\
         \cline{2-5}
         {Preference over combined-phase 1D} & \multicolumn{4}{c}{4.9$\sigma$}\\
         {Preference over separated-phase 1D} & \multicolumn{2}{c}{3.6$\sigma$} & \multicolumn{2}{c}{-}\\
         \hline
    \end{tabular}
    \caption{Retrieved Fe abundances with HyDRA-2D for each side of the atmosphere and with the 1D retrievals for each phase range (see Figure~\ref{fig:fe_constraints}). We also show the model preference for each retrieval in $\sigma$. We multiply the overall evidences of the separated-phase retrievals to allow for comparisons with the combined-phase retrievals. We see no preference for HyDRA-2D over a separated-phase 1D retrieval for the $\phi = 0.02 \, - \, 0.04$ range.}
    \label{tab:fe_sigma_constraints}
\end{table*}

\begin{figure*}
\centering
	\includegraphics[width=\textwidth,trim={0cm 0cm 0cm 0},clip]{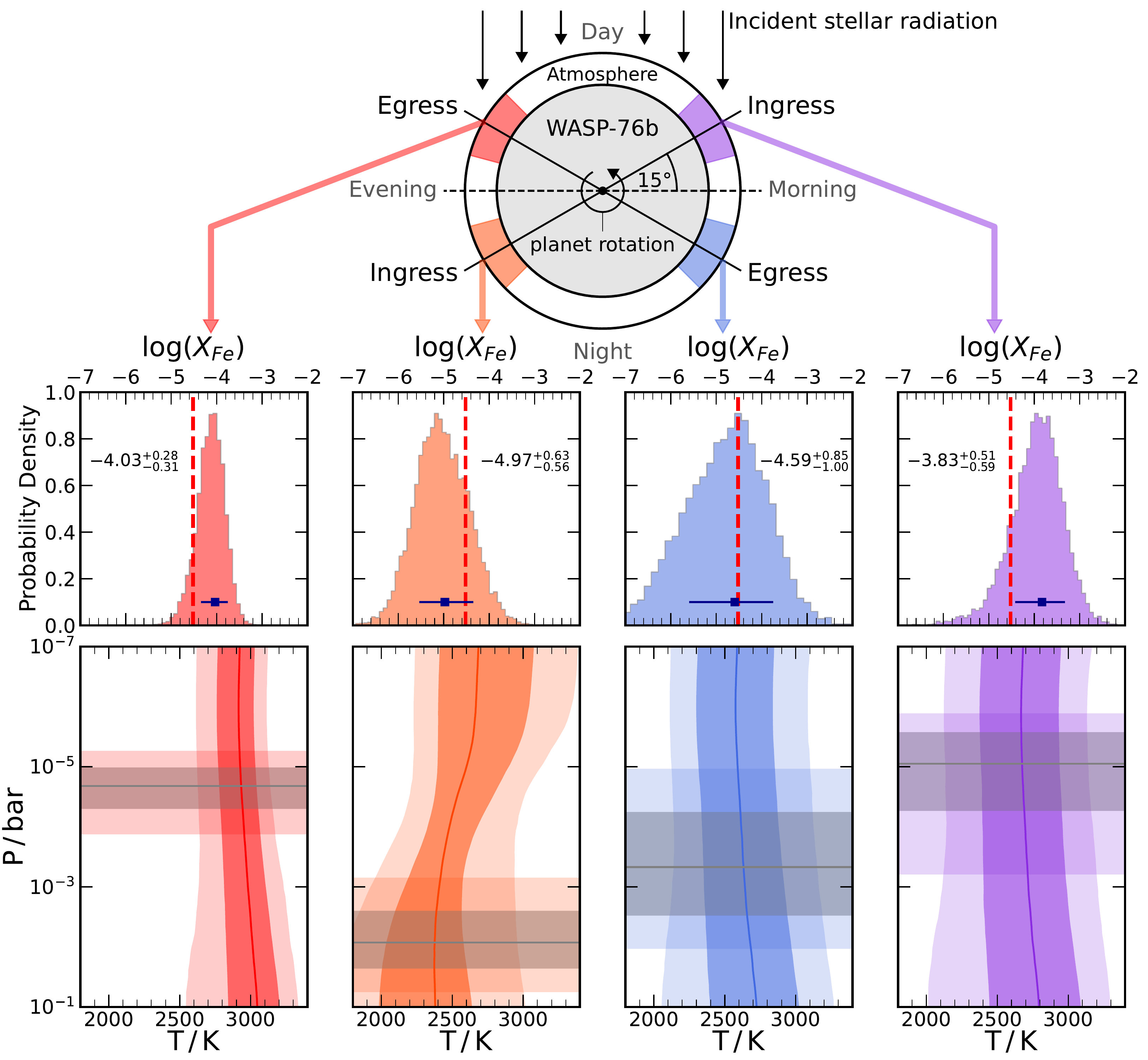}
    \caption{Constraints on the Fe, temperature profile and opacity deck for WASP-76~b with HyDRA-2D. At the top of the figure we show a schematic of the region of the atmosphere probed by the retrievals, with the corresponding constraints for each region highlighted below. For the Fe histograms the blue error bar indicates the 1$\sigma$ confidence interval and the red dashed line shows the solar value of Fe (see Figure~\ref{fig:fe_constraints}). The bottom panels show the 1$\sigma$ and 2$\sigma$ confidence interval for the P-T profile in the darker and lighter shades respectively, with the median value given by the solid curve. Similarly, we highlight the constraints on the opacity deck and its 1$\sigma$ and 2$\sigma$ confidence interval in the darker and lighter grey. Note that this is the uncertainty in the pressure of the top of the opacity deck, and the observed spectrum is not sensitive to higher pressures as the opacity deck is optically thick.}
\label{fig:fe_t_constraints}
\end{figure*}

\begin{figure}
\centering
	\includegraphics[width=0.49\textwidth,trim={0.0cm 0cm 0cm 0},clip]{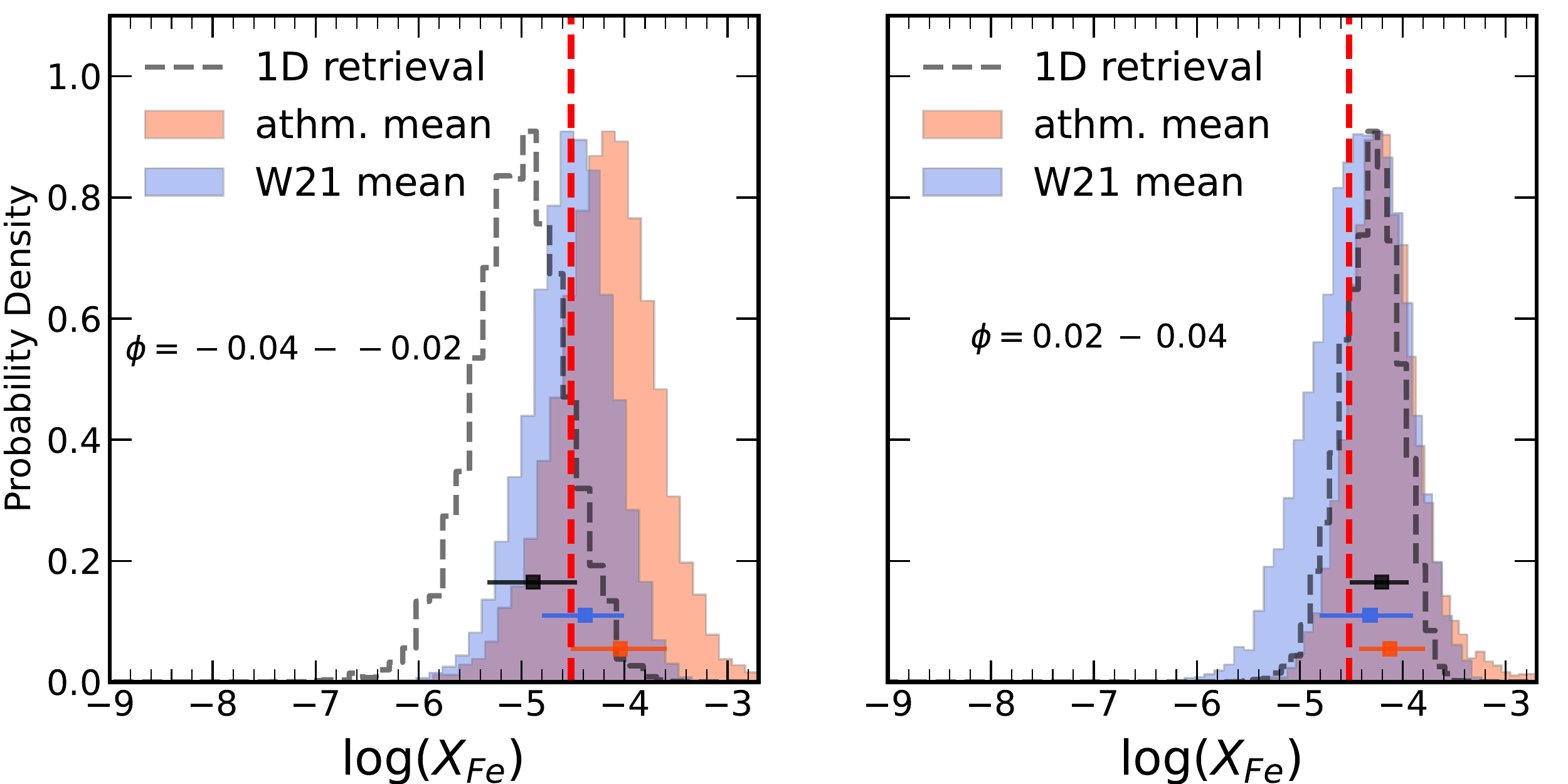}
    \caption{Averaged posterior distribution of HyDRA-2D over the $\phi = -0.04 \,\text{-}\, -0.02$ and $\phi = 0.02 \,\text{-}\, 0.04$ ranges. We calculate an arithmetic mean (shown in red) and a temperature-weighted geometric mean based on the prescription in \citet{welbanks2022} (shown in blue). We also show the results from the separated-phase 1D retrievals with the dashed black line. The red dashed line indicates the solar value of Fe.}
\label{fig:fe_means}
\end{figure}

\begin{figure}
\centering
	\includegraphics[width=0.49\textwidth,trim={0.0cm 0cm 0cm 0},clip]{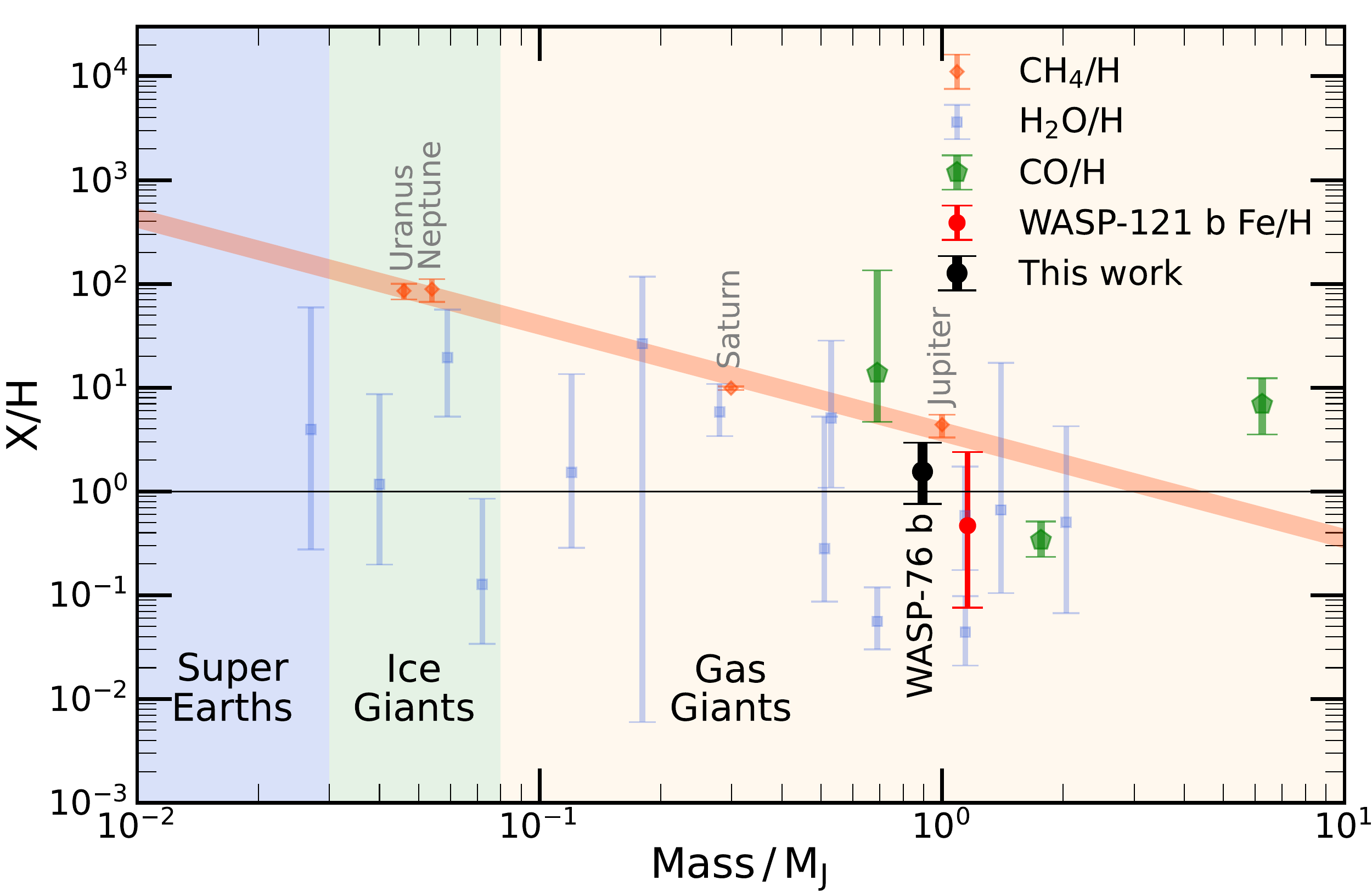}
    \caption{Mass-metallicity relation for known exoplanets and the Solar System, corrected for the stellar metallicity of each system. The constraint from this work is shown in black and uses our most precise value of $\log(\mathrm{Fe}) = -4.03^{+0.28}_{-0.31}$, which arises from the evening side for the last quarter of the transit. The orange region indicates the Solar System trend of decreasing metallicity with mass. We also show the Fe abundance measurement in the ultra-hot Jupiter WASP-121~b \citep{gibson2022} in red, and H$_2$O/H (blue) and CO/H (green) constraints for a range of exoplanets obtained with low-resolution and high-resolution observations respectively \citep{welbanks2019, guillot2022}.}
\label{fig:mass_met}
\end{figure}

To compare our results with HyDRA-2D we also perform a spatially-homogeneous retrieval across the combined $\phi = -0.04 \,\text{-}\, -0.02$ and $\phi = 0.02 \,\text{-}\, 0.04$ ranges, which we refer to as the combined-phase 1D-retrieval. We also perform two additional spatially-homogeneous retrievals for the $\phi = -0.04 \,\text{-}\, -0.02$ and $\phi = 0.02 \,\text{-}\, 0.04$ ranges separately (referred to as the separated-phase 1D-retrievals). Such spatially-homogeneous retrievals are well established and have been used to analyse HST as well as ground-based high-resolution observations \citep[e.g.,][]{brogi2019, gandhi2019_hydrah, line2021}. We assume uniform Fe abundance across the whole terminator for these 1D-retrievals, with a single pressure-temperature profile for the atmosphere and uniform opacity deck and haze. Hence for each retrieval we have 13 free parameters, including $\Delta K_\mathrm{p}$, $\Delta V_\mathrm{sys}$ and $\delta V_\mathrm{wind}$. When applying the spectral broadening to these models we take the full rotational kernel for both halves of the atmosphere. Further details on the retrieval setup for the 1D retrievals can be found in \citet{gandhi2018} and \citet{gandhi2019_hydrah}.

\section{Results and Discussion}

In this section we discuss the results of the HyDRA-2D retrievals on the WASP-76~b observations. We show the constraints for Fe in Figure~\ref{fig:fe_constraints} and the constraints on the temperature profile and opacity deck in Figure~\ref{fig:fe_t_constraints}. These indicate that the median values and their errors vary significantly between the evening and morning, and between the retrievals across the $\phi = -0.04 \,\text{-}\, -0.02$ and $\phi = 0.02 \,\text{-}\, 0.04$ phase ranges. Table~\ref{tab:fe_sigma_constraints} also shows that HyDRA-2D offers a better statistical fit the observations for the $\phi = -0.04 \,\text{-}\, -0.02$ range over traditional spatially-homogeneous approaches. This therefore highlights the need for such modelling capabilities in order to effectively constrain atmospheric parameters and resolve spatial differences between the two sides. On the other hand, we find that for the last quarter of the transit the evening side dominates the signal, with little contribution from the morning side, and hence the 1D retrieval shows almost equal evidence. We discuss the constraints for each of the various parameters in detail below.

\subsection{Comparison to spatially-homogeneous models}

We compare our constraints on Fe with more traditional 1D modelling approaches, where homogeneous terminator properties are assumed. We find that our spatially-homogeneous combined-phase 1D retrieval obtains strong constraints on the Fe abundance, with $\log(X_\mathrm{Fe}) = -4.33^{+0.26}_{-0.27}$ (see Table~\ref{tab:fe_sigma_constraints}). When we perform separate retrievals for each of the phase ranges, we find that the abundance constraints now show some differences. The separate-phase 1D retrieval retrieves $\log(X_\mathrm{Fe}) = -4.89^{+0.43}_{-0.45}$ for the $\phi = -0.04 \,\text{-}\, -0.02$ range. On the other hand, the signal from the last quarter of the transit closely matches the Fe abundance constraint from the combined-phase 1D retrieval. Therefore, the combined-phase retrieval constrains the most significant contribution to the Fe signal, which comes from the latter quarter, consistent with previous studies \citep{ehrenreich2020, kesseli2021}. Given the differences in the constrained abundance, it is therefore unsurprising that the separated-phase 1D retrievals are statistically preferred by 3.8$\sigma$. This indicates that the overall atmospheric structure is not as well modelled by assuming homogeneous physical properties across the transit.

The HyDRA-2D constraints go further by exploring the variation in Fe over the terminator for each of the phase ranges. The evening sides of the spatially-resolved retrievals show remarkably similar constraints to the separated-phase 1D retrievals (see Figure~\ref{fig:fe_constraints} and Table~\ref{tab:fe_sigma_constraints}). For both phase ranges the evening side is thus the dominant side of the planet's Fe signal, which remains true for the $\phi = -0.04 \,\text{-}\, -0.02$ range even though the evening side's abundance is constrained to be lower than for the morning. This is because the spectral contribution resulting from a lower abundance is offset by the much deeper opacity deck for the evening side (see section~\ref{sec:results_cl} and Figure~\ref{fig:fe_t_constraints}), thus maintaining a stronger contribution to the overall signal. The statistical preference for HyDRA-2D and the separated-phase 1D retrievals over the combined-phase retrieval is driven primarily by differences in the temperature and the wind speeds (see sections~\ref{sec:results_T} and \ref{sec:results_winds}), with a weaker effect caused by the chemical inhomogeneities of Fe in the atmosphere. This is consistent with previous work by \citet{beltz2021} comparing high resolution observations in emission with general circulation models.

We also calculate terminator-averaged abundances from HyDRA-2D and compare these to our separated-phase 1D retrievals for each of the phase ranges we study, as shown in Figure~\ref{fig:fe_means}. We do this to investigate the potential biases which may arise by assuming spatial homogeneity. We calculate the average abundance of the terminator from HyDRA-2D through two methods. Firstly, we use the arithmetic mean of Fe across the two terminators. Our second method employs a more sophisticated temperature-weighted geometric mean, which has been shown to provide a better prescription for averaged terminator abundances for low-resolution observations \citep{welbanks2022}. This calculation does implicitly assume that the cross section between the two sides does not vary significantly, as we do not weight each side by the overall opacity for each half of the terminator. However, we can justify this given that we do not expect significant variation in the opacity over the range of the temperature constraints. 

The averaged values are consistent with each other for both phase ranges given the relatively little difference in abundance between the two sides. Our 1D retrieval is in fact most closely matching the evening side abundance from HyDRA-2D (see Figure~\ref{fig:fe_constraints}). This is likely due to the fact that with the high resolution R=140,000 observations of ESPRESSO we obtain separate signals from each side of the atmosphere due to their difference in velocity. Hence the signals from each side of the atmosphere are not completely blended together, as would be the case for low resolution observations, and thus for our high-resolution observations the spatially-unresolved retrievals converge to the stronger signal. The last quarter of the transit ($\phi = 0.02 \,\text{-}\, 0.04$) shows similar constraints for both methods of averaging the abundance as the constraint from the morning is relatively weak and thus dominated by the evening side.

\subsection{Variation in Fe}

The Fe is most strongly constrained from HyDRA-2D for the evening side of the atmosphere for the $\phi = 0.02 \,\text{-}\, 0.04$ range (see Figure~\ref{fig:fe_constraints}). We see a tight well-constrained abundance of $\log(X_\mathrm{Fe}) = -4.03^{+0.28}_{-0.31}$, slightly above the solar value of $\log(X_\mathrm{Fe}) = -4.52$ \citep{asplund2009}, but consistent with the metallicity of the host star, [Fe/H] = 0.366. However, our highest abundance estimate for Fe in fact comes from the morning of the $\phi = -0.04 \,\text{-}\, -0.02$ range, with $\log(X_\mathrm{Fe}) = -3.83^{+0.51}_{-0.59}$. Although, it should be noted that this estimate has a higher uncertainty. Both of these regions of the atmosphere with the highest abundance estimates probe more of the day side of the atmosphere, i.e., that which is being irradiated due to the geometry of the transit (see Figure~\ref{fig:fe_t_constraints}). Hence it is unsurprising that we obtain the tightest constraints for these regions, as the day side is expected to have the strongest Fe signal given the hotter temperatures, as discussed in section~\ref{sec:results_T} below.

Our morning side constraint for the $\phi = 0.02 \,\text{-}\, 0.04$ range has the lowest abundance and the weakest constraint, at $\log(X_\mathrm{Fe}) = -4.59^{+0.85}_{-1.0}$, as the evening side dominates the overall signal. The stellar limb darkening also acts against the signal from the morning side for the latter half of the transit, as this side occults more of the dimmer limbs of the star. Given the wider uncertainty for the morning side, it is unsurprising that the 1D retrieval shows a similar evidence and constrains an Fe abundance which closely matches the evening. Further observations of the transit of WASP-76~b may constrain the Fe signal from the morning more definitively, particularly with high resolution (R$\sim$140,000) observations which are able to resolve the leading and trailing limbs. 

We find that the abundance constraints for the sides probing more of the night side, i.e. $\phi = -0.04 \,\text{-}\, -0.02$ evening side and $\phi = 0.02 \,\text{-}\, 0.04$ morning side, have lower median values by $\sim$0.8~dex, indicating that some Fe rain-out may be occurring across the terminator, as suggested by \citet{ehrenreich2020} for the morning side. This rain-out is also consistent with recent work on cloud formation on UHJs by \citet{gao2021} that showed that the Fe abundance could potentially decrease by $\sim1$~dex due to it condensing. However, we note that the chemical abundance differences between the various regions of the atmosphere are generally quite small and broadly consistent with each other, and differences in the temperature and wind speeds have a higher impact on the overall spectrum.

\subsubsection{Mass-metallicity relation}

Figure~\ref{fig:mass_met} shows our most precise abundance estimate for Fe, that of the evening side for the last quarter of the transit, along with previous measurements of various species in Solar System planets and other exoplanets. As chemical models of the atmospheres of such hot objects have shown \citep[e.g.,][]{visscher2010}, gaseous iron is expected to be the dominant proportion of the Fe, and thus a good estimate of the total Fe content of the atmosphere. The Fe measurement of an exoplanet atmosphere also provides a direct comparison against the stellar abundance \citep[e.g.,][]{thiabaud2015}. Our retrieved value lies in between the stellar metallicity and the trend from the Solar System planets. In addition to WASP-76~b, we also show the Fe abundance measured for WASP-121~b \citep{gibson2020}, a UHJ with an equilibrium temperature of $\sim$2400~K, and the CO abundance constraints from other high resolution studies \citep{gandhi2019_hydrah, pelletier2021, line2021}. These show good agreement with our result, as well as the previous work of \citet{welbanks2019} which found species other than oxygen at stellar- or super-stellar abundances. 

The Fe abundance is of particular importance given that it can be used to determine the refractory-to-volatile abundance ratio. This provides a unique window into the formation and migration scenarios \citep{lothringer2021}. As the Fe accreted onto the planet must be in the form of solids, it allows us to break the degeneracy often seen with volatile species, where the carbon and oxygen bearing species may be accreted as either solids or gases \citep[e.g.,][]{oberg2011, mordasini2016}. Further observations of UHJs and constraints on their refractory species will help us to discern the trends and allow us to compare with the volatile abundances measured from observations in the infrared.

\subsection{Variation in the thermal structure}\label{sec:results_T}

Our retrieved temperature and its constraints do show variation between the evening and morning sides as well as with phase, as shown in Figure~\ref{fig:fe_t_constraints}. These show that the temperature is best constrained for the evening side towards the end of the transit, where we retrieve a hot photosphere of $T_\mathrm{0.1mb} = 2950^{+111}_{-156}$~K. On the other hand, we obtain a lower median temperature with a wider uncertainty of $T_\mathrm{0.1mb} = 2615^{+266}_{-275}$~K for the morning side for the $\phi = 0.02 \,\text{-}\, 0.04$ phase range. The constraints for both phase ranges show that the parts of the atmosphere that are irradiated show the highest temperatures, in line with expectations from general circulation models \citep{wardenier2021, savel2022}. For all of our temperature constraints the profiles also appear largely isothermal, and generally show a widening uncertainty at pressures higher than the opacity deck where the observations are unable to probe. GCMs indicate thermal inversions for the upper photosphere for the day side, although these inversions are generally consistent with the retrieved uncertainty in our temperature profiles for all cases except the evening side for the $\phi = 0.02 \,\text{-}\, 0.04$ range, where our error is more tightly constrained given the higher signal-to-noise.

We compare our temperature constraints to previous analyses and other observations of the primary transit of WASP-76~b. High resolution observations using the CARMENES spectrograph constrained a higher temperature of $\sim$2700-3700~K range, consistent with our temperature profiles \citep{landman2021}. Recent analyses with the ESPRESSO and HARPS observations in the optical detected a temperature of $\sim$3300-3400~K \citep{seidel2019, seidel2021}, which is higher than our temperature for even the hottest part of the atmosphere, namely the evening side for the last quarter of the transit. However, \citet{vonessen2020} used low resolution HST observations in the optical and found temperature constraints of $2300^{+412}_{-392}$~K, while \citet{edwards2020} analysed HST WFC3 observations in the infrared and obtained $2231^{+265}_{-283}$~K. \citet{fu2021} combined the WFC3 and Spitzer observations and found that the best fit temperature was 2000~K from forward models. These low resolution constraints point to a temperature that is generally lower than we observe, but still consistent at $\sim$1.5$\sigma$ with our constraint for the morning side 0.1~mbar temperature for both phase ranges. The differences may be reconciled given that the high resolution observations in our work are probing a higher altitude than the low resolution observations and thus may obtain a higher temperature if there is a stratosphere present, as indicated from day side analysis of HST observations \citep{fu2021}. In future, combining the high-resolution and low-resolution observations may help to constrain the thermal profile to a greater extent \citep[e.g.,][]{brogi2019, gandhi2019_hydrah} and shed light on the potential differences seen in the constrained temperature of the terminator.

\begin{figure}
\centering
	\includegraphics[width=0.49\textwidth,trim={0.0cm 0cm 0cm 0},clip]{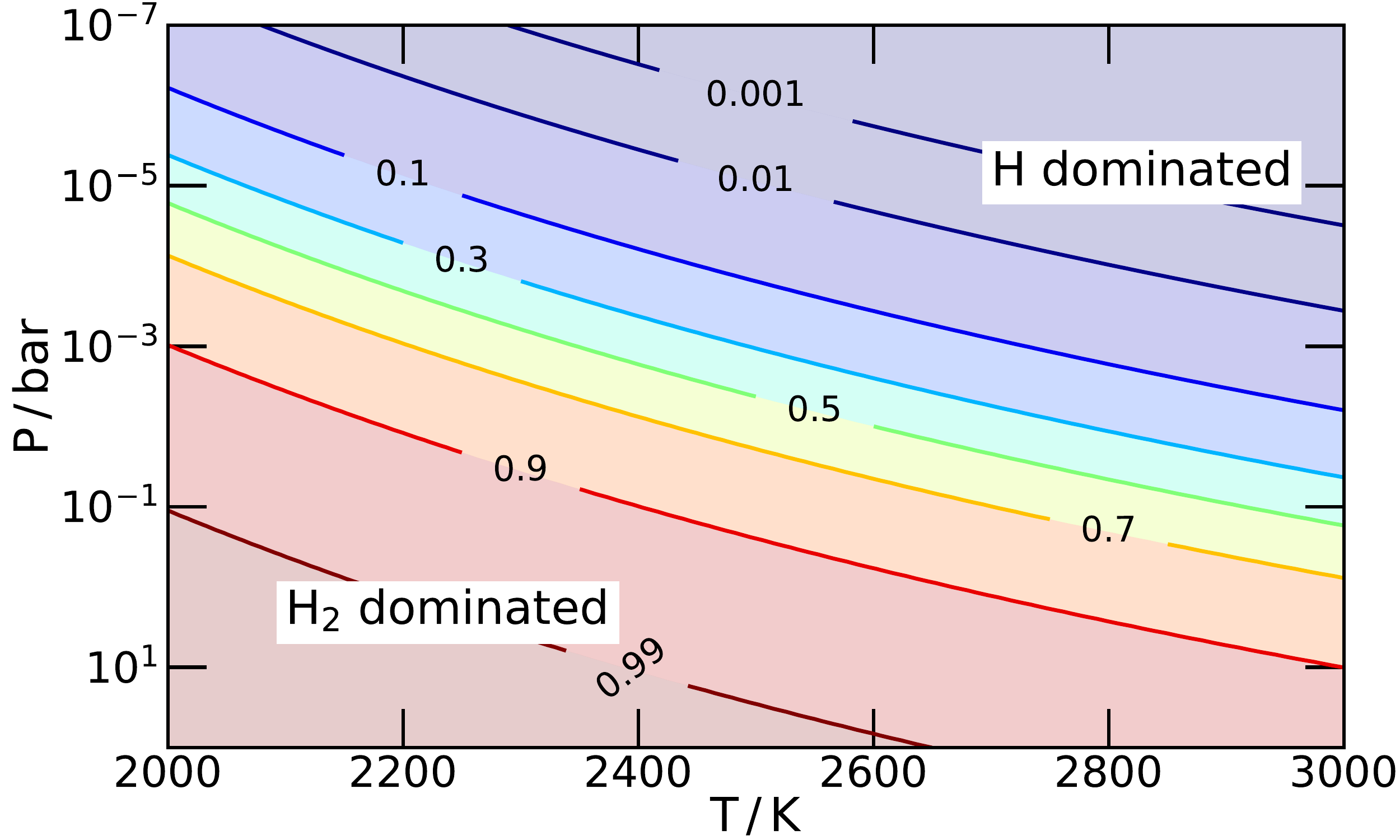}
    \caption{Fraction of H$_2$ to (H+H$_2$) for typical photospheric pressures and temperatures for ultra-hot Jupiters. These were derived using the analytic prescription in \citet{parmentier2018} (see also \citet{gandhi2020_h-}). The lowest pressures with the highest temperatures have the greatest degree of H$_2$ dissociation, resulting in atmospheres which are H rich.}
\label{fig:h2_dissoc}
\end{figure}

We have performed retrievals assuming an H$_2$/He rich and an H/He rich atmosphere, and find that the H/He rich case (where the H$_2$ is dissociated) offers a substantially better fit to the observations. This is because the atmospheric composition is intricately linked to the overall constrained temperature. Such a dissociated atmosphere, consisting primarily of H and He, would be a lower mean molecular weight over an H$_2$ and He rich atmosphere by a factor of almost two. Therefore, the scale height, given by $H = k_b T/\mu g$, is almost double for an H/He rich atmosphere over an H$_2$/He rich atmosphere. Modelling of circulation on UHJs has shown that the majority of the day side atmosphere is H$_2$-dissociated at pressures $\sim$0.1~bar \citep{bell2018, tan2019}. In the upper atmosphere ($\sim10^{-4}$~bar) where our observations are more sensitive, we expect a significant majority of the hydrogen in the atmosphere to be thermally dissociated, as shown in Figure~\ref{fig:h2_dissoc}. Therefore, our retrievals with an H$_2$/He rich atmosphere constrained unphysically high temperatures of $\gtrsim3800$~K in order to match the strength of the features in the observations. Furthermore, we also find that the H$_2$/He rich retrievals were statistically disfavoured, and thus find evidence for an atmosphere that is dissociated. The recombination of H into H$_2$ on the morning side of the last quarter of the transit may also be a potential reason for its lack of significant signal compared to the evening side, as its scale height would be decreased by a factor of $\sim$2 over the evening in addition to any difference in temperature and Fe abundance. 

We do observe a weak thermal inversion for the evening side in the $\phi = -0.04 \,\text{-}\, -0.02$ range. This may be driven by a potential thermal inversion from the day side remaining present at the terminator, and detectable given the deeper opacity deck for this part of the atmosphere (see section~\ref{sec:results_cl}). Numerous strong optical absorbers are capable of generating thermal inversions in ultra-hot Jupiters \citep[e.g.,][]{lothringer2018, gandhi2019_inv}. We note however that the retrieved inversion may also be caused by the recombination of H$_2$ in the deep atmosphere. At high pressures, the H$_2$ is present at significant abundance as shown in Figure~\ref{fig:h2_dissoc}. This increases the mean molecular weight and thus reduces the scale height. Hence our retrieval with the H/He rich atmosphere may act to reduce the apparent scale height in the lower atmosphere by lowering the temperature at higher pressures, and thus appear as a thermal inversion. We tested our temperature profile parametrisation was not the root cause of this inversion by testing various pressures where the temperature was set instead of 0.1~mbar, and also varying the lower prior on $P_1$ and $P_2$, but these did not impact our results. We do note however that the inversion is weak ($\sim300$~K) and consistent with an isotherm as these high-resolution terminator observations are only indirectly sensitive to the temperature profile from its effect on the scale height. We performed an additional retrieval which did not allow thermal inversions in the atmosphere and observed no significant change to our parameter estimates, further indicating that the inversion not too significant. 

\begin{figure*}
\centering
    \includegraphics[width=0.49\textwidth,trim={0.0cm 0cm 0cm 0},clip]{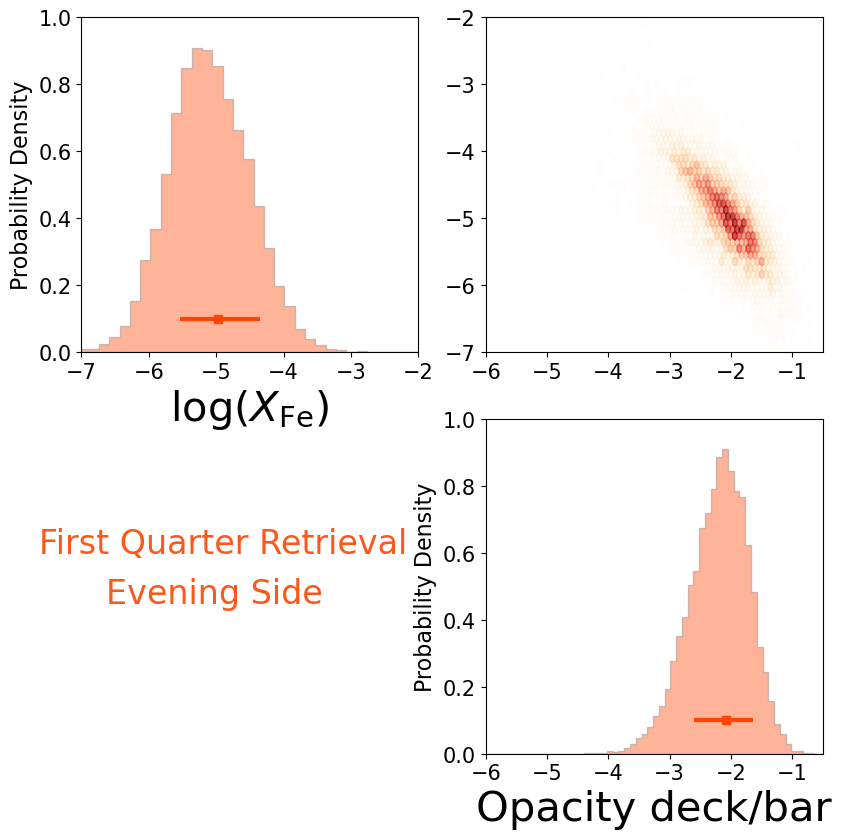}
	\includegraphics[width=0.49\textwidth,trim={0.0cm 0cm 0cm 0},clip]{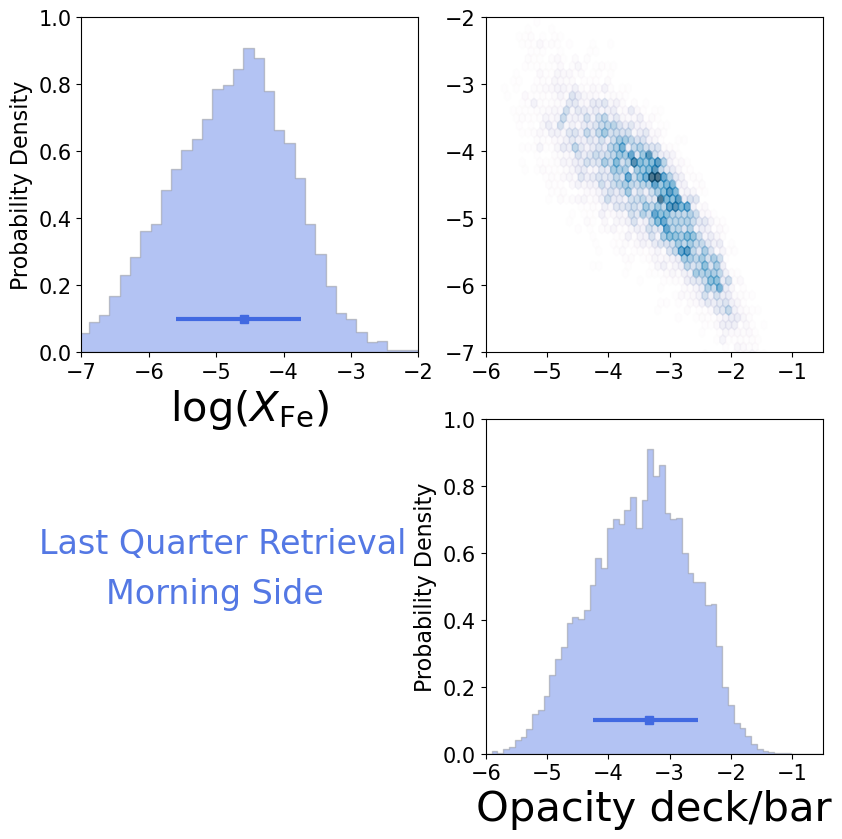}
    \caption{Posterior distribution of our retrievals of WASP-76~b highlighting the degeneracy between Fe abundance and opacity deck. The left panels in red show the degeneracy for the evening side of the first quarter retrieval ($\phi = -0.04 \,\text{-}\, -0.02$), and the right hand panels shows the constraints for the morning side for the last quarter ($\phi = 0.02 \,\text{-}\, 0.04$).}
\label{fig:fe_cloud}
\end{figure*}

\subsection{Opacity deck}\label{sec:results_cl}

The constraints for the opacity deck are given in Figure~\ref{fig:fe_t_constraints}. The range in pressures constrained for the top of the cloud deck in Figure~\ref{fig:fe_t_constraints} represents the uncertainty, and the spectrum is not sensitive to pressures below this as the opacity deck is assumed to be optically thick. The retrievals show that there is a high opacity deck/continuum present for the sides of the atmosphere which are probing most strongly the day side of the atmosphere, and our signal for Fe is probing the upper atmosphere above the opacity deck. The source of this opacity likely arises from species such as H- and/or other atomic species which may be prevalent in the atmosphere, as condensation of cloud species is unlikely on the day side for such a hot day side. This is backed up by the numerous species which have also been detected in the atmosphere with significant optical opacity \citep{tabernero2021, kesseli2022}. We note that species such as TiO or VO also have strong optical opacity and are capable of leading to thermal inversions \citep{hubeny2003, fortney2008, piette2020}, but there is no evidence for these species from the high-resolution observations \citep{tabernero2021}. On the other hand, our results for the non-irradiated regions of the atmosphere show a deeper opacity deck, indicating that the night sides may be relatively free of any strong optical absorbers. Cloud formation modelling has shown that the night side should possess significant cloud opacity due to silicate clouds \citep{gao2021}. Such clouds form at pressures of 10$^{-2}$-10$^{-3}$~bar depending on the temperature, consistent with our opacity decks for the night side of $\log(P_\mathrm{cl}/\mathrm{bar}) = -2.08^{+0.44}_{-0.53}$ (evening side, $\phi = -0.04 \,\text{-}\, -0.02$) and $\log(P_\mathrm{cl}/\mathrm{bar}) = -3.33^{+0.80}_{-0.91}$ (morning side, $\phi = 0.02 \,\text{-}\, 0.04$). Recent work has also shown that patchy clouds may be present on the night side of UHJs but that they only weakly affect terminator observations \citep{komacek2022}.

\subsubsection{Abundance-opacity deck degeneracy}

Our constrained Fe abundance shows a degeneracy with the pressure level of the opacity deck, as shown in Figure~\ref{fig:fe_cloud}. This shows that a higher altitude opacity deck requires a higher abundance in the atmosphere, and arises because the column depth of a given species increases to match the observed feature strength when the cloud deck altitude is raised \citep[e.g.,][]{betremieux2017, welbanks2019_degen}. Recent work on characterising atmospheres with cloud/opacity decks at high resolution \citep{gandhi2020_clouds} indicated that abundances of chemical species are still constrainable even in the presence of high altitude opacity decks. Our R=140,000 observations with EPSRESSO confirm this is the case and we obtain relatively tight constraints for Fe despite our limited spectral coverage of just $\sim$0.4~$\mu$m.

We find that the Fe abundance retrieved for the morning side for $\phi = 0.02 \,\text{-}\, 0.04$ has a wider uncertainty due to the abundance-opacity deck degeneracy. \citet{ehrenreich2020} proposed that Fe rains out of the atmosphere and thus we are unable to observe it for the morning side later in the transit. On the other hand, GCMs have shown that the lack of Fe signal may be due to the presence of a cloud deck obscuring the signal from Fe \citep{savel2022}. Our retrievals indicate a slight preference for the latter, but the constraints are weak due to its weaker signal compared to the much hotter evening side. We also see no preference for our spatially-resolved retrieval over a 1D spatially-homogeneous retrieval for the $\phi = 0.02 \,\text{-}\, 0.04$ range (see Table~\ref{tab:fe_sigma_constraints}), indicating that the Fe abundance of the morning may not be significantly different to that from the evening.

\subsection{Constraints on winds}\label{sec:results_winds}

\begin{figure}
\centering
\begin{overpic}[width=0.49\textwidth]{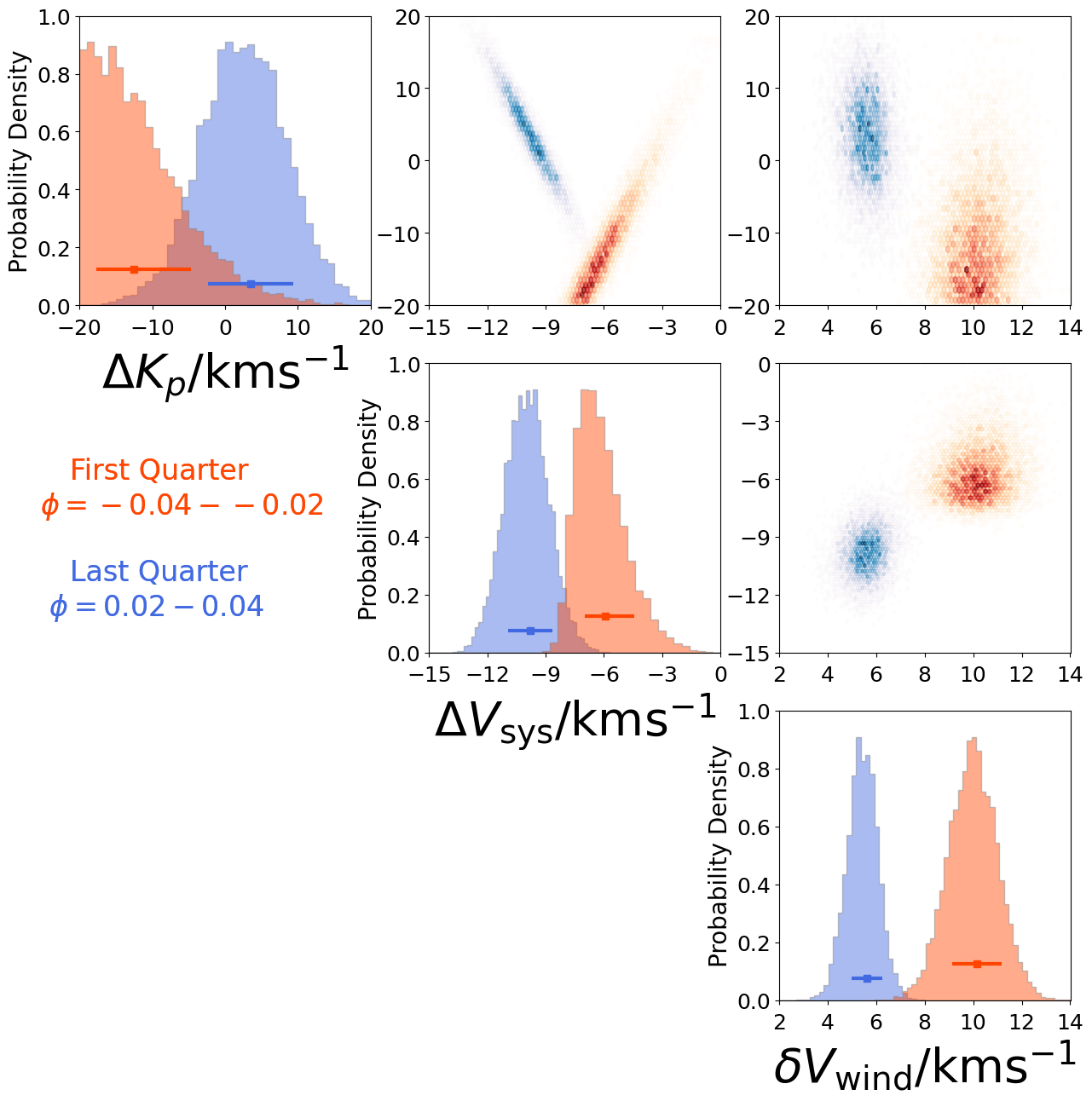}\large
\put (-5,17) {\def\arraystretch{1.3}
\small
\begin{tabular}{l|{c}r}
\textbf{Param.}              & \textbf{First Quarter} & \textbf{Last Quarter} \\
\hline
$\Delta K_\mathrm{p}$ & $-12.5\substack{+7.8 \\ -5.2}$ & $3.5\substack{+5.9 \\ -5.9}$ \\
$\Delta V_\mathrm{sys}$ &$-5.9\substack{+1.5 \\ -1.1}$ & $-9.8\substack{+1.2 \\ -1.1}$ \\
$\delta V_\mathrm{wind}$ & $10.2\substack{+1.0 \\ -1.0}$ & $5.62\substack{+0.62 \\ -0.64}$ \\
\end{tabular}
}
\end{overpic}
    \caption{Posterior distributions of $\Delta K_\mathrm{p}$ and the wind parameters, $\Delta V_\mathrm{sys}$ and $\delta V_\mathrm{wind}$, for each HyDRA-2D retrieval of WASP-76~b. The red posteriors show the constraints on the $\phi = -0.04 \,\text{-}\, -0.02$ range and the blue posteriors show the $\phi = 0.02 \,\text{-}\, 0.04$ range. We also provide the median and $\pm1\sigma$ values for each retrieval in the table on the bottom left.}
\label{fig:kp_vsys_vwind}
\end{figure}

\begin{figure}
\centering
	\includegraphics[width=0.49\textwidth,trim={0.0cm 0cm 0cm 0},clip]{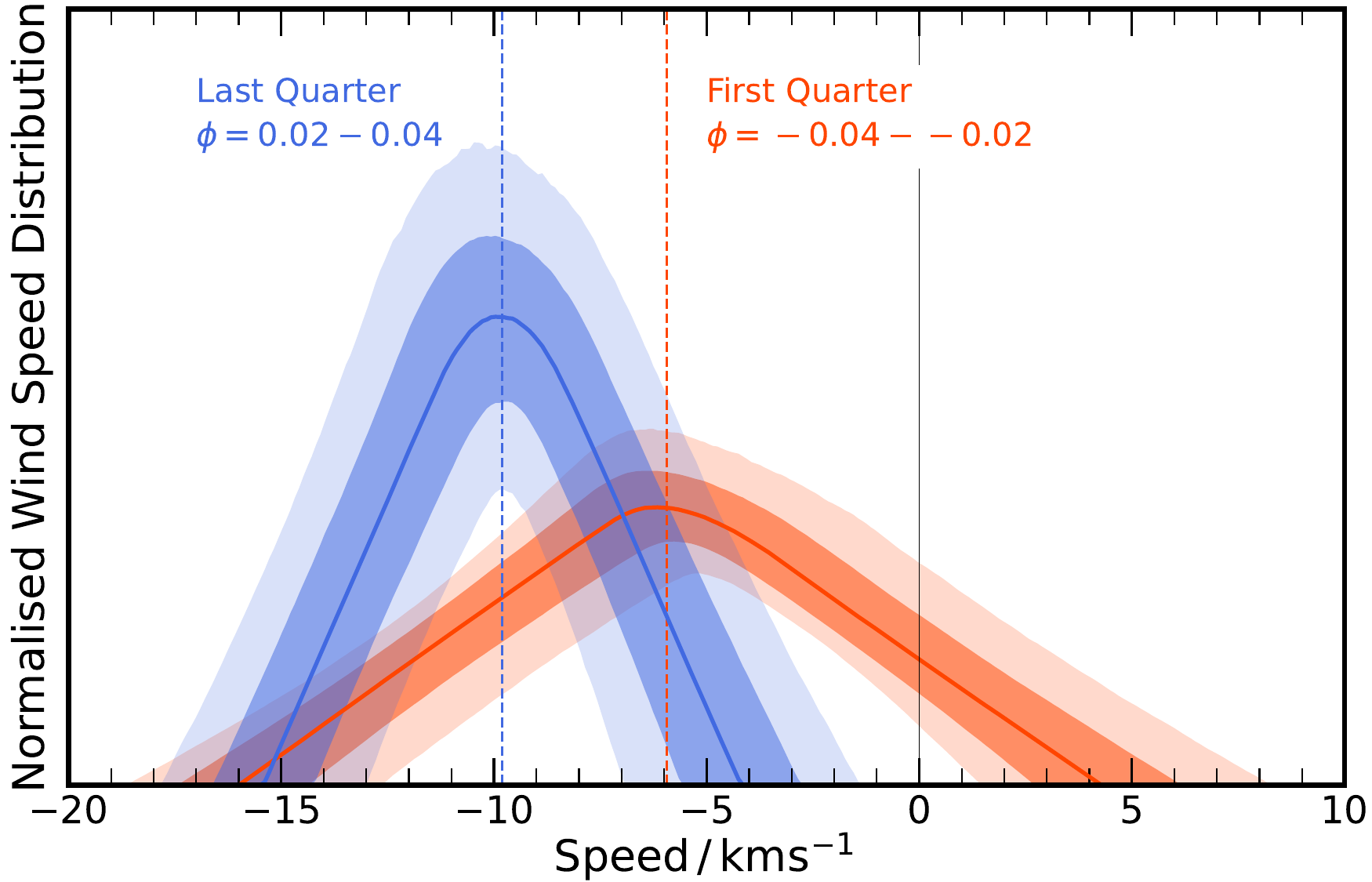}
    \caption{Normalised wind speed distribution as derived from the $\Delta V_\mathrm{sys}$ and $\delta V_\mathrm{wind}$ parameters given in Figure~\ref{fig:kp_vsys_vwind}. This is the day-night wind broadening kernel applied to the model as shown in Figure~\ref{fig:broaden_schem}. The constraints on the $\phi = -0.04 \,\text{-}\, -0.02$ range are shown in red and the constraints on the $\phi = 0.02 \,\text{-}\, 0.04$ range in blue. The dashed lines indicate the median values of $\Delta V_\mathrm{sys}$ from our retrieval. The negative values indicate a day-to-night wind, which results in a net blue-shift of the overall spectrum.}
\label{fig:wind_speeds}
\end{figure}

The constraints on the wind parameters from HyDRA-2D are shown in Figure~\ref{fig:kp_vsys_vwind} and the corresponding constraints on the wind speed distributions are plotted in Figure~\ref{fig:wind_speeds}. In our case the wind speed is given by $\Delta V_\mathrm{sys}$ as we have shifted the spectra to the rest frame of the planet during our analysis, and the uncertainty in the known value of $V_\mathrm{sys}$ is negligible. The $\Delta V_\mathrm{sys}$ values are dependent on the orbital period and transit time centre, but by propagating the uncertainties reported in \citet{ehrenreich2020}, we find that $\Delta V_\mathrm{sys}$ changes by at most $\sim$0.5~km/s. By using other recently published precise orbital values \citep{Kokori2022, ivshina2022}, we also find consistent results to within the uncertainties and $\sim$1 km/s. Therefore, any red- or blue-shift of the spectrum indicates the presence of a wind and cannot be explained by variations in orbital parameters. 

For both phase ranges we see a strong day-night wind with a significant blue shift to the overall spectrum, consistent with previous observations and analyses \citep{ehrenreich2020, kesseli2021}. This is a result of the day-night wind which travels toward the observer at the terminator. For the first quarter of the transit we retrieve a wind speed of $-5.9^{+1.5}_{-1.1}$~km/s, with the negative sign indicating a blue-shift (i.e. a wind travelling from day to night side). This value is consistent with previously derived wind speeds from the detection of Na in the upper atmosphere of WASP-76~b \citep{seidel2021}. However, we constrain a stronger wind speed of $-9.8^{+1.2}_{-1.1}$~km/s for the $\phi = 0.02 \,\text{-}\, 0.04$ range, but Na is not as strongly detected in this phase range \citep{kesseli2022} and thus could explain why the wind speeds are not in as good an agreement.

As well as the difference in the wind speeds, we also see a difference in the $\delta V_\mathrm{wind}$ parameter, which measures the full-width half maximum of the spread in wind speed. We constrain a smaller $\delta V_\mathrm{wind}$ for the last quarter of the transit, as shown in Figure~\ref{fig:wind_speeds}. The differences in the parameters during the transit could potentially be due to a combination of a more dominant day-night wind and a weaker east-west jet. As the signal from the $\phi = 0.02 \,\text{-}\, 0.04$ range is dominated by the evening side, any jet present would add to the overall profile and result in a greater blue shift of the signal, resulting in a lower $\Delta V_\mathrm{sys}$. In addition, the spread in the $\delta V_\mathrm{wind}$ would also be lower as the morning side contributes very little to the overall signal. On the other hand, the $\phi = -0.04 \,\text{-}\, -0.02$ range has a significant contribution from both sides, and may therefore result in a lower overall wind speed and greater spread as any contribution from an east-west jet will be red-shifted for the morning. In our constraints we do retrieve a wind profile that has a small red-shifted contribution for the first quarter of the transit. A day-night wind of $\sim$6~km/s combined with a jet of $\sim$4~km/s could thus explain the relative shift in $\Delta V_\mathrm{sys}$ and spread in $\delta V_\mathrm{wind}$ between the first and last quarter of the transit. We have also carried out retrievals with an east-west jet instead of a day-night wind and did not see strong evidence for such a case, thus any jet present would be weak compared to the day-night wind. This is consistent with the GCMs from \citet{wardenier2021}, who showed that such a weak-drag model best describes the observations. 

\begin{figure*}
\centering
    \includegraphics[width=0.49\textwidth,trim={0.0cm 0cm 0cm 0},clip]{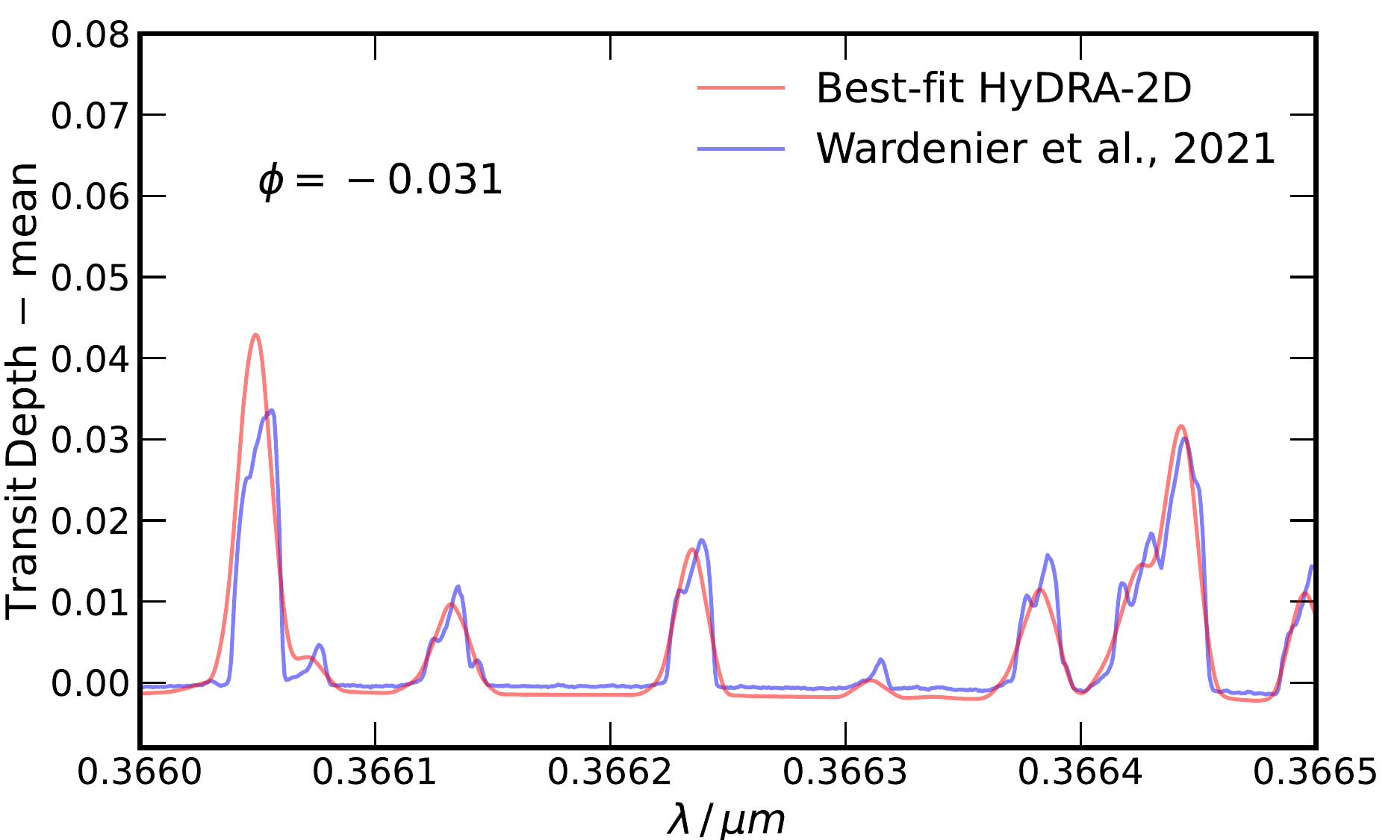}
	\includegraphics[width=0.49\textwidth,trim={0.0cm 0cm 0cm 0},clip]{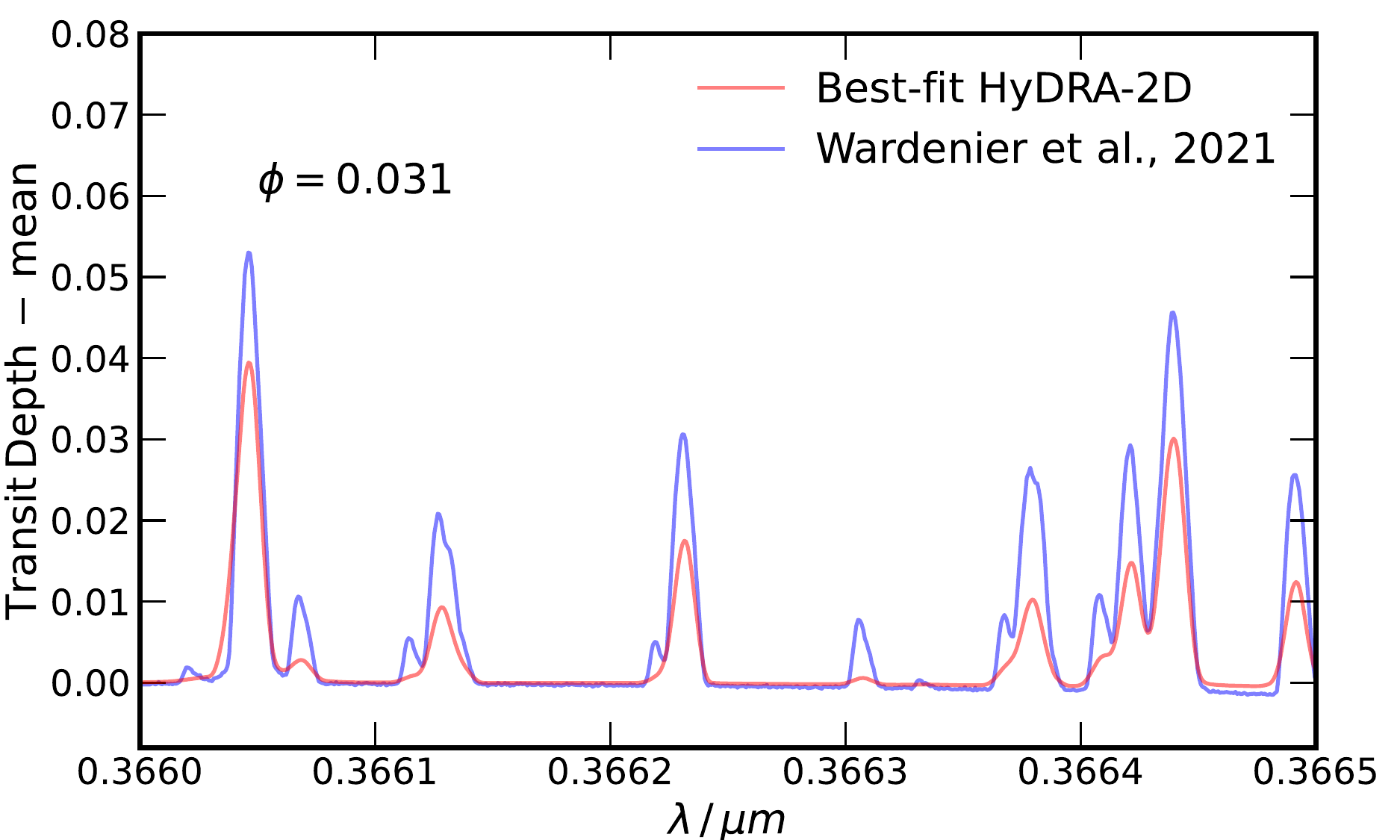}
    \caption{Comparison of HyDRA-2D to general circulation models of WASP-76~b. The highest likelihood model from the retrieval of the $\phi = -0.04 \,\text{-}\, -0.02$ range was used to generate a transit spectrum as shown in the left panel (in red), and the right panel shows the the corresponding best fit model for the $\phi = 0.02 \,\text{-}\, 0.04$ range. The general circulation model assumes an atmosphere with weak drag, and is calculated through Monte Carlo radiative transfer \citep[see][for further details]{wardenier2021}.}
\label{fig:gcm_compare}
\end{figure*}

On the other hand, the higher opacity decks constrained for both the morning and evening terminators for the $\phi = 0.02 \,\text{-}\, 0.04$ range may also result in such a difference in wind parameters between the first and last quarter of the transit. If the observations in the latter half of the transit probe higher altitudes, there may be more of a blue shift in the spectrum as the day-night wind is expected to be highest at lower pressures. This would also reduce the overall spread in the wind speed, $\delta V_\mathrm{wind}$, as the range in pressures probed reduces. Hence it is currently unclear which of these two potential causes of the difference in wind speed is dominant, and further observations of WASP-76~b at high resolution will help to shed light on the various dynamical processes occurring in the atmospheres of UHJs. 

We note that there is a strong degeneracy between $\Delta K_\mathrm{p}$ and $\Delta V_\mathrm{sys}$ in Figure~\ref{fig:kp_vsys_vwind}, as expected given the limited phase range of the observations which would act to break this. This degeneracy results in the two $\Delta K_\mathrm{p}$-$\Delta V_\mathrm{sys}$ distributions for the first and last quarter, which tend towards each other at lower $K_\mathrm{p}$ values. However, the true $K_\mathrm{p}$ value is constrained to within $\sim$1 km/s \citep{ehrenreich2020}, and so we believe the $\Delta V_\mathrm{sys}$ offset between the first and last quarter of the transit is a result of the observations and not uncertainty in the $K_\mathrm{p}$. The $\Delta K_\mathrm{p}$ is consistent with expected value of 0 for the $\phi = 0.02 \,\text{-}\, 0.04$ range, but the $\phi = -0.04 \,\text{-}\, -0.02$ range does indicate a lower $\Delta K_\mathrm{p}$, albeit still consistent with 0 at $\lesssim$2$\sigma$. The deviation from the expected value is be due to the signal changing its velocity with phase for the first part of the transit, as seen in previous studies with other datasets and GCMs \citep[e.g.,][]{kesseli2021, wardenier2021}. This phase dependence is not modelled in our retrievals due to the complexity of including multi-dimensional phase-dependent radiative transfer into our calculations, and thus results in a slight change in $\Delta K_\mathrm{p}$ to compensate.

\subsection{Comparison to GCMs}

We compare the results from HyDRA-2D to general circulation models of WASP-76~b in Figure~\ref{fig:gcm_compare}. These show the asymmetric weak-drag Monte Carlo radiative transfer model of \citet{wardenier2021} (see ``modification 1'' in their Figures 12 and 13) at phases of $\phi = -0.031$ and $\phi = 0.031$, along with the best fit retrieval model for the $\phi = -0.04 \,\text{-}\, -0.02$ and $\phi = 0.02 \,\text{-}\, 0.04$ range for a small region near $\sim$0.366~$\mu$m. For both of the best fit models we have removed the opacity deck as Fe is expected to be the dominant source of opacity in this spectral range, and we do not expect significant cloud condensation or hazes given the high temperature. We have also mean subtracted the datasets to remove the effect of varying planetary radius and/or reference pressure, as these dependencies are removed in our retrieval in the high-pass filter as part of the analysis. 

Figure~\ref{fig:gcm_compare} shows that for the first quarter of the transit, the two models agree well. This is crucial as 3-dimensional models have shown better fits to high resolution observations \citep{flowers2019, beltz2021}, and thus HyDRA-2D is able to capture the relevant physics to fit the dynamics and variability in the observations. The line positions agree well and have similar strengths between our model and the GCM, indicating that the constrained wind speeds, abundances and temperatures are in line with expectations. We do however find that our retrieved best-fit models are somewhat smoother than the predictions made from GCMs. This difference is because we model an atmosphere with a latitudinally homogeneous day-night wind contribution, but the GCM with weak drag includes latitude-dependent effects, depth dependent wind speeds as well as a temperature structure that varies along the line of sight. These additional effects result in additional structure to the GCM spectra. We do also see some small differences in the strengths of some of the Fe lines, but the overall agreement between the retrieval and GCM is encouraging.

The latter half of the transit shows that the retrieved features are weaker than the GCM predictions. The GCM indicates that the strength of the Fe signal should increase, but the retrieved best fit model from HyDRA-2D for $\phi = 0.031$ does not increase in strength over the $\phi = -0.04 \,\text{-}\, -0.02$ range. A potential factor driving this may be the temperature profile. The retrieval constrains temperatures nearer $\sim$3000~K for the evening side, but the GCM shows some regions of the atmosphere reaching temperatures nearer to $\sim$3500~K. This difference results in larger spectral features in the GCM spectra by $\sim15$\% due to the changing scale heights of the atmosphere. There will also be an additional and potentially stronger effect of varying line strengths with temperature, and the weaker Fe lines being further reduced in the retrieval due to their strong temperature dependence as shown in Figure~\ref{fig:gcm_compare}. We note that as the retrieval is a quantitative fit to the data, the GCM may not be completely capturing the true state of the atmosphere and overestimating the temperature. This temperature difference between the retrieval and GCM may be be driven by various effects such as a super-solar Fe abundance, other optical absorbers as well as disequilibrium chemistry in the atmosphere. Further observations of the transit of WASP-76~b will help shed more light on these differences and help improve our 3D modelling of such hot planets. 

\subsection{Robustness tests}

To verify the robustness of our results we performed numerous checks with our model. We carried out retrievals with a scale factor which rescales the model spectrum to account for uncertainties in the measured value of the radius and mass and any non-hydrostatic effects which may be present in the upper atmosphere. Such a parameter has often been included in previous high resolution retrievals \citep[e.g.,][]{gandhi2019_hydrah, gibson2020, line2021}. The scale factor did show some degeneracy with the temperature as both act to increase/decrease the overall scale height of the atmosphere, but there were no differences in the retrieved parameters and the overall relative evidences remained similar. We also performed separate retrievals on the $\phi = 0.02 \,\text{-}\, 0.035$ and $\phi = 0.025 \,\text{-}\, 0.04$ range, and $\phi = -0.035 \,\text{-}\, -0.02$ and $\phi = -0.04 \,\text{-}\, -0.025$ ranges. The top pressure that we modelled our atmosphere out to was also varied from 10$^{-7}$~bar to 10$^{-8}$~bar. As well as these tests, we performed retrievals with additional free parameters for the radius and $\log(g)$ of the planet. None of the retrievals showed any significant difference to any parameter, and our retrievals with $\log(g)$ constrained estimates consistent with the known values \citep{ehrenreich2020}. We were unable to constrain the planetary radius to any significant degree and thus simply recovered the prior given that the high-pass filtering of the data analysis removes the radial dependence of the spectrum. We also performed retrievals varying the number of live points with MultiNest and did not see any significant change in the convergence of any of the parameters.

We also performed additional retrievals parametrising the temperature profile as an isotherm and with no super-Rayleigh haze prescription, and found no significant changes to our constraints on the photospheric temperature, Fe abundance, opacity deck or wind parameters. The evidences for the isothermal retrievals was also similar, with the largest difference coming from the HyDRA-2D retrieval for the first quarter given that a weak thermal inversion was constrained for the evening side. However, this was $<2\sigma$ deviant from the retrievals with the fully flexible P-T profile. These tests indicate that the observations are more sensitive to the overall temperature than to the gradients in temperature with height, in line with expectations for primary eclipse observations.

The only change we saw in the retrieved constraints was when we included the very bluest end of the data between 0.38-0.4~$\mu$m. For these tests we retrieved a higher temperature for the atmosphere and a negative velocity shift in $\Delta K_\mathrm{p}$. Given the very strong Fe opacity, these wavelengths probe regions of the atmosphere $\lesssim$10$^{-7}$~bar where non-local thermodynamic effects are likely to be considerable, and our assumptions of ideal gas and hydrostatic equilibrium break down. Hence it is unsurprising that our retrieval obtains a lower overall evidence for this wavelength region. We also varied the wavelength range we use for the redder end of the observations, with retrievals in the 0.4-0.65~$\mu$m and 0.4-0.7~$\mu$m range as well as our fiducial 0.4-0.8~$\mu$m. These showed no change in the retrieved constraints for any parameter, despite some telluric absorption from O$_2$ and H$_2$O at wavelengths $\gtrsim0.7$~$\mu$m. This confirms that the spectral cleaning steps were able to effectively remove the tellurics in this wavelength range (see section~\ref{sec:data_analysis}).

\section{Conclusions}

We have explored variations in Fe abundance, temperature profile and opacity/haze deck in the atmosphere of the ultra-hot Jupiter WASP-76~b through a new spatially-resolved high resolution retrieval of the terminator. This was motivated by recent analyses on ESPRESSO and HARPS observations of the primary transit which indicated the presence of an asymmetric signal in Fe during the transit. With our new retrieval model, HyDRA-2D, we explored the variability and constrained a day-night wind in the photosphere. We find that Fe has the tightest constraints in the observations on the trailing (evening) limb of the last quarter of the transit ($\phi = 0.02 \,\text{-}\, 0.04$), with an abundance of $\log(X_\mathrm{Fe}) = -4.03^{+0.28}_{-0.31}$. The abundance estimates of Fe show that the atmosphere is consistent with the stellar-metallicity and the CH$_4$/H measurement of Jupiter, as well as with previous constraints of Fe on the UHJ WASP-121~b \citep{gibson2022}.

Our HyDRA-2D models are statistically preferred by 4.9$\sigma$ over more traditional spatially-homogeneous and combined-phase models. The different regions of the atmosphere probed are distinguishable given the high precision and resolution of the observations made with the ESPRESSO spectrograph. The statistical preference is driven primarily by differences in the temperature and winds, with a weaker effect also caused by the chemical abundance differences between the different regions of the atmosphere. This is consistent with previous 3D modelling of high resolution emission spectra in the infrared \citep{beltz2021}. Furthermore, for our separated-phase retrievals, we find that HyDRA-2D is preferred by 3.6$\sigma$ over a spatially-homogeneous approach for the $\phi = -0.04 \,\text{-}\, -0.02$ range. Hence adopting a spatially-homogeneous model is likely to miss signals from some regions of the atmosphere, particularly when both sides have high signal-to-noise and there is significant asymmetry in the physical parameters. On the other hand, we find that our 1D-retrieval for the $\phi = 0.02 \,\text{-}\, 0.04$ range has nearly identical evidence to the spatially-resolved HyDRA-2D model, as a result of the evening side dominating over the morning side for the latter quarter of the transit.

Generally, we observe Fe abundances which are consistent between the different regions of the terminator, but we do find that the Fe abundance is highest where the region of the planet probed is being irradiated, namely the morning side for $\phi = -0.04 \,\text{-}\, -0.02$ and the evening side for $\phi = 0.02 \,\text{-}\, 0.04$. We also find that our spatially-homogeneous retrievals for the first and last quarter of the transit converge towards the constraints of the evening side. This suggests that the evening side is the dominant source of the Fe signal. This remains true for the first quarter of the transit despite the higher constrained abundance for the morning side, as this higher abundance is offset by the higher altitude opacity deck in the atmosphere, resulting in a weaker overall contribution.

We do not constrain Fe as significantly on the morning side for the $\phi = 0.02 \,\text{-}\, 0.04$ range, with the retrieved abundance, $\log(X_\mathrm{Fe}) = -4.59^{+0.85}_{-1.0}$, having a wider uncertainty than for any other part of the atmosphere. \citet{ehrenreich2020} proposed that the Fe signal is most prominent on the evening, and the lack of significant signal from the morning side may be due to Fe condensing out of the atmosphere. On the other hand, \citet{savel2022} proposed that the signal was diminished due to a high altitude cloud deck. Our retrievals show that either scenario can fit the observations given that there is a degeneracy between abundance and the opacity deck, with a slight preference for a high abundance and opacity/cloud deck. The degeneracy is only partial as recent work has shown that abundance constraints from high-resolution observations are attainable even with high altitude opacity/cloud decks \citep{gandhi2020_clouds}, and we confirm this with observational data. However, our Fe uncertainty for the morning for the $\phi = 0.02 \,\text{-}\, 0.04$ range is much wider than for other regions of the atmosphere because the signal from the evening side dominates the overall spectrum, as suggested from GCMs \citep{wardenier2021}.

Our constraints on the temperature profile indicate an atmosphere that is between $2950^{+111}_{-156}$~K ($\phi = 0.02 \,\text{-}\, 0.04$, evening side) and $2615^{+266}_{-275}$~K ($\phi = 0.02 \,\text{-}\, 0.04$, morning side) at 0.1~mbar. We generally find a temperature that is hottest for the parts of the atmosphere that are irradiated, i.e. the evening side for the last quarter of the transit and the morning side for the first quarter. Our temperature profiles do not significantly constrain the temperature gradient of the atmosphere, but this is expected given that we do not directly detect photons from the planet in primary transit thus making constraints on the profile more difficult. We do however see a strong dependence of the temperature on thermal dissociation, a dissociated atmosphere consisting of H and He was statistically preferred and gave more physically plausible temperatures over an undissociated atmosphere of H$_2$ and He. This is because an undissociated atmosphere has a smaller scale height and thus the temperature must substantially increase in order to compensate and match the features in the spectrum. Hence we infer the presence of H$_2$ dissociation in the atmosphere of WASP-76~b, consistent with predictions made for ultra-hot Jupiters \citep[e.g.,][]{parmentier2018}.

We constrain a much stronger day-night wind at the end of the transit at $-9.8^{+1.2}_{-1.1}$~km/s, where the negative sign denotes that the wind is blue-shifted and hence travelling towards the observer at the terminator. For the last quarter of the transit the signal is dominated by the evening side. On the other hand, the first quarter of the transit includes strong contributions from both sides of the atmosphere, and we find that the wind speed is reduced to $-5.9^{+1.5}_{-1.1}$~km/s. The FWHM of the wind speed distribution, $\delta V_\mathrm{wind}$, is also more spread for the $\phi = -0.04 \,\text{-}\, -0.02$ range at $10.2^{+1.0}_{-1.0}$~km/s than for the $\phi = 0.02 \,\text{-}\, 0.04$ range at $5.62^{+0.62}_{-0.64}$~km/s. These constraints could indicate the presence of a weak equatorial jet in the photosphere, as this would induce a stronger blue shift of the evening side and combine with the day-night wind to drive the overall wind speed to higher values for the $\phi = 0.02 \,\text{-}\, 0.04$ range. However, the stronger wind speed and lower $\delta V_\mathrm{wind}$ could also be a result of the higher altitudes probed in this phase range, given that the day-night wind is expected to be stronger at higher altitudes \citep[e.g.,][]{wardenier2021}.

This work is a significant step into multi-dimensional spatially-resolved retrieval analyses of exoplanetary atmospheres. Key to these retrievals is incorporating the 3-dimensional dynamical processes of GCMs with the flexibility and computational efficiency required to explore millions of models over a wide ranging parameter space. Potential future work could make the wind profiles more realistic with more complex physics \citep[e.g.,][]{seidel2020}, or implement a varying temperature profile along the line of sight for each side of the atmosphere \citep{nixon2022, dobbs-dixon2022}, more akin to the thermal profiles from GCMs. Another potential improvement may be to use the likelihood method of \citet{gibson2020}. Such models are also ideally suited to simultaneous analyses combining high- and low-resolution observations \citep{gandhi2019_hydrah}, and will play a key role as we begin characterising exoplanets with the next generation of facilities.

\section*{Acknowledgements}

SG is grateful to Leiden Observatory at Leiden University for the award of the Oort Fellowship. This work was performed using the compute resources from the Academic Leiden Interdisciplinary Cluster Environment (ALICE) provided by Leiden University. We also utilise the Avon HPC cluster managed by the Scientific Computing Research Technology Platform (SCRTP) at the University of Warwick. IS acknowledges funding from the European Research Council (ERC) under the European Union’s Horizon 2020 research and innovation program under grant agreement No 694513. MB acknowledges support from from the UK Science and Technology Facilities Council (STFC) research grant ST/T000406/1. JPW sincerely acknowledges support from the Wolfson Harrison UK Research Council Physics Scholarship and the Science and Technology Facilities Council (STFC). LW thanks support provided by NASA through the NASA Hubble Fellowship grant \#HST-HF2-51496.001-A awarded by the Space Telescope Science Institute, which is operated by the Association of Universities for Research in Astronomy, Inc., for NASA, under contract NAS5-26555. ABS acknowledges support from the Heising-Simons Foundation. We also thank the anonymous referee for a careful review of our manuscript.

%%%%%%%%%%%%%%%%%%%%%%%%%%%%%%%%%%%%%%%%%%%%%%%%%%
\section*{Data Availability}

The models underlying this article will be shared on reasonable request to the corresponding author.



\bibliographystyle{mnras}
\bibliography{refs} % if your bibtex file is called example.bib

%\begin{thebibliography}{99}
%\end{thebibliography}

%%%%%%%%%%%%%%%%%%%%%%%%%%%%%%%%%%%%%%%%%%%%%%%%%%

%%%%%%%%%%%%%%%%% APPENDICES %%%%%%%%%%%%%%%%%%%%%

%\appendix


%If you want to present additional material which would interrupt the flow of the main paper, it can be placed in an Appendix which appears after the list of references.

%%%%%%%%%%%%%%%%%%%%%%%%%%%%%%%%%%%%%%%%%%%%%%%%%%

% Don't change these lines
\bsp	% typesetting comment
\label{lastpage}
\end{document}